\documentclass[preprint,pteplogo]{ptephy}

\preprintnumber{KUNS-2513} 

\usepackage{ulem}
\newcommand{\Slash}[1]{{\ooalign{\hfil/\hfil\crcr$#1$}}}
\newcommand{\ppp}{{\pi^0\pi^+\pi^-}}
\newcommand{\p}{{3\pi^0}}
\newcommand{\tr}{{\rm tr}}

\usepackage{graphicx}	
\usepackage{color}	
\begin{document}

\title{
The $\eta$ decay into 3$\pi$ in asymmetric nuclear medium}

\author{Shuntaro Sakai and Teiji Kunihiro}

\address{Department of Physics, Kyoto University,
Kitashirakawa-Oiwakecho, Sakyo-ku, Kyoto 606-8502, Japan\\
\email{s.sakai@ruby.scphys.kyoto-u.ac.jp}}
\begin{abstract}%
We explore how the $\eta-\pi^0$ mixing angle and the $\eta$ meson decay
into $\pi^{+}\pi^{-}\pi^0$ and 3$\pi^{0}$ 
are modified in the  nuclear medium on the basis of the in-medium chiral
 effective theory with varying isospin asymmetry $\alpha$, where
$\alpha\equiv \delta\rho/\rho$ with $\delta \rho=\rho_n-\rho_p$ and
$\rho=\rho_n+\rho_p$.
We find  that
the larger the isospin asymmetry $\delta \rho$ and 
the smaller the total density $\rho$,
the more enhanced  the mixing angle.
We show that the decay width in the nuclear medium has 
an additional density dependence that cannot be renormalized into 
that of the mixing angle: The additional term originates from the vertex
 proportional to a low-energy constant $c_1$,
which only comes into play in the nuclear medium but not in the free space.
It turns out that the resultant density effect on the decay widths
 overwhelms that coming from the isospin asymmetry, and
 the higher the $\rho$,
the more enhanced the decay widths; the width for the
 $\pi^{+}\pi^{-}\pi^0$ decay is enhanced by a factor of two to three at
 the normal density $\rho_0$ with a minor increase due to  $\delta
 \rho$, while that for the 3$\pi^0$ decay shows only a small increase of
 around 10\% even at $\rho_0$.
We mention the possible relevance of the partial restoration of
chiral symmetry to  the unexpected density effect on the decay widths
in the nuclear medium.
\end{abstract}
\subjectindex{xxxx, xxx}
\maketitle
\section{Introduction\label{intro}}
Quantitative understanding of $\eta$ meson decay into three
$\pi$'s is a long standing problem in hadron physics
\cite{Sutherland1966,Bell1968,Weinberg1975,Roiesnel1981,Gasser1985,Abdou2003,Bijnens2007,Schneider2011,Lanz2013}.
The decay process is prohibited by the $G$ parity conservation
with isospin symmetry being taken for granted, and the electromagnetic
correction is found to vanish in the leading order \cite{Sutherland1966}.
The origin of the decay is attributed to the isospin symmetry breaking
inherent in quantum chromodynamics (QCD), the small current-quark mass
difference between $u$ and $d$ quarks\footnote{Regarding this $\eta$ decay
process, S.~Weinberg dealt with this process as  the U$_A$(1)
problem \cite{Weinberg1975} related to the $\eta$ mass or $\pi^0$,
$\eta$, and $\eta'$ mixing properties and various attempts were suggested to
explain the experimental data; see, e.g.,
Refs.~\cite{Kogut1975,Raby1976,Kawarabayashi1980}.}.
Due to the isospin symmetry breaking, the observed $\eta$ and
$\pi^0$ do not correspond to the eigenstate of the flavor SU(3) but to
their mixed state;
\begin{align}
 \begin{pmatrix}
  \eta\\
  \pi^0
 \end{pmatrix}=
 \begin{pmatrix}
  \cos\theta&-\sin\theta\\
  \sin\theta&\cos\theta 
 \end{pmatrix}
 \begin{pmatrix}
  \eta_8\\
  \pi_3
 \end{pmatrix},\label{defma}
\end{align}
where we denote the mass eigenstate as $\eta$ and $\pi^0$, and the
flavor SU(3) eigenstate as $\eta_8$ and $\pi_3$.
The angle $\theta$ is called the $\eta-\pi^0$ mixing angle.
The $\pi_3$ component of the $\eta$ meson enables the meson to decay
into 3$\pi$, showing that the mixing angle plays an essential role in
the decay.
However, an analysis based on the current algebra shows a
large discrepancy with the experimental result \cite{Bell1968}.
Recent theoretical development has revealed the significance of the
final-state interaction between pions in addition to the $u$ and $d$
quark mass difference for the quantitative account of the
experimental
data\footnote{In Refs.~\cite{Schechter1971,Hadnull1974}, a
fairly good agreement is obtained in the linear sigma model analysis.
The linear sigma model contains the explicit iso-singlet sigma meson
degree of freedom, and the sigma mesons have some relationship with the
s-wave $\pi\pi$ correlation.
The recent analysis of the experimental data supports the existence of
the $\sigma$ pole in the $\pi\pi$ channel ($I=0$) \cite{Beringer2012}.
The importance of the two-$\pi$ correlation in the
$\eta-3\pi$ system is discussed in Ref.~\cite{Roiesnel1981}, as
mentioned in the text.
Here, we note that it is reported in Ref.~\cite{Abdou2003} that
the inclusion of the scalar mesons in a different manner hardly affects
the $\eta$ decay width.}.
A phenomenological inclusion of the effect of the final-state
interaction \cite{Roiesnel1981} shows a good
agreement with the observed decay width and it is shown that the higher
order contribution of chiral perturbation theory to the decay process
that contains the effect of the final-state interaction is quite
large \cite{Gasser1985,Bijnens2007,Schneider2011}.

The modification of the hadron properties in the environment, which is
characterized by temperature, baryonic density,
electromagnetic field, and so on,
is an interesting topic of hadron physics (see, e.g.,
Refs.~\cite{Hatsuda1994,Brown1996}).
Furthermore, the effects of isospin asymmetry in a nuclear medium on
hadron properties are being investigated in various systems, including
the pion$-$nucleus system \cite{Friedman2007,Hayano2010}, the few-body
system of hyper-nuclei \cite{Akaishi2000}, and the equation of
state of nuclear matter \cite{Shinmura2002,Ueda2013}.
In this paper, we show that the $\eta$ decay into three pions provides 
yet another example revealing the interesting effects caused by
the isospin-asymmetry of the nuclear medium.

As for the in-medium properties of the $\eta$
meson, some intensive searches for $\eta$-bound states in
a nucleus has been made (for a recent review, e.g., \cite{Kelkar2013}), 
where the focus is put on the $\eta$-optical potential and/or a possible
mass shift in the nuclear medium.

In this paper, we study the effect of the asymmetric nuclear
medium, focusing on the three-$\pi$ decay of $\eta$ in the
isospin-asymmetric nuclear medium, where the isospin-breaking background
field is present in addition to the different $u$ and $d$ quark
masses\footnote{An early version of the present work was presented in
Ref.~\cite{Sakai:2014xqa}}.
We shall show that the isospin asymmetry in the nuclear
medium increases the mixing of the $\eta$ and $\pi^0$ mesons and 
thereby leads to  an enhancement of  the decay  rates of the $\eta$
meson to three pions.
We shall also find that the (isosymmetric) total baryon density
unexpectedly causes an  enhancement of the decay width, which is found
to be associated with the phenomenon of the partial restoration of
chiral symmetry in the nuclear medium.

This paper is organized as follows.
In Sect.~\ref{how}, we introduce the model Lagrangian and explain
the calculation of the $\eta-\pi^0$ mixing angle and the decay
amplitude of the $\eta$ into three $\pi$ in the asymmetric nuclear
medium in chiral effective field theory.
We evaluate the $\eta-\pi^0$ mixing angle in the
asymmetric nuclear medium in Sect.~\ref{secma}.
Then we discuss the decay amplitude of the $\eta$ meson decay into
3$\pi$ in the isospin-asymmetric nuclear medium with numerical results
in Sect.~\ref{sece3p}.
A brief summary and concluding remarks are presented in Sect.~\ref{conc}.
In the Appendix, we present the explicit forms of 
the meson$-$baryon vertices derived from the chiral Lagrangian and used in
the calculation in the text.

\section{Preliminaries\label{how}}
To investigate the $\eta$ meson decay into three $\pi$ in the nuclear
medium, we apply the chiral effective field theory in the nuclear medium.
A full account of the in-medium chiral perturbation may be seen in
Refs.~\cite{Meissner2002,Kaiser2002}.
The basic degrees of freedom are the flavor-octet pseudoscalar
mesons and baryons. Then the chiral Lagrangian needed for our
calculation reads \cite{Gasser1985b,Gasser1988}
\begin{align}
 \mathcal{L}=&\mathcal{L}^{(2)}_{\pi\pi}+\mathcal{L}^{(4)}_{\pi\pi}+\mathcal{L}_{\pi
 N}^{(1)}+\mathcal{L}_{\pi N}^{(2)},\label{lag}\\
 \mathcal{L}_{\pi\pi}^{(2)}=&\frac{f^2}{4}\left< D_\mu
					 UD^\mu U^\dagger+\chi U^\dagger+U\chi^\dagger\right>,\\
 \mathcal{L}_{\pi\pi}^{(4)}=&L_1\left<D_\mu U(D^\mu
					U)^\dagger\right>^2+L_2\left<D_\mu
					U(D_\nu 
					U)^\dagger\right>\left<D^\mu
					U(D^\nu U)^\dagger\right>\notag\\
&+L_3\left< D_\mu U(D^\mu
						    U)^\dagger D_\nu
						    U(D^\nu U)^\dagger
 \right>+L_4\left<D_\mu U(D^\mu U)^\dagger\right>\left<\chi
    U^\dagger+U\chi^\dagger\right>\notag\\
&+L_5\left<D_\mu U(D^\mu
    U)^\dagger(\chi U^\dagger+U\chi^\dagger)\right>+L_6\left<\chi
    U^\dagger+U\chi^\dagger\right>^2+L_7\left<\chi
    U^\dagger-U\chi^\dagger\right>^2\notag\\
&+L_8\left<U\chi^\dagger
    U\chi^\dagger +\chi U^\dagger\chi
    U^\dagger\right>^2-iL_9\left<f_{\mu\nu}^RD^\mu U(D^\nu
    U)^\dagger+f^L_{\mu\nu}(D^\mu U)^\dagger D^\nu
 U\right>\notag\\
&+L_{10}\left<Uf^L_{\mu\nu}U^\dagger f^{\mu\nu}_R\right>+H_1\left<f^R_{\mu\nu}f_R^{\mu\nu}+f^L_{\mu\nu}f_L^{\mu\nu}\right>+H_2\left<\chi\chi^\dagger\right>,\\
 \mathcal{L}_{\pi
  N}^{(1)}=&\left<\bar{B}\left(i\Slash{D}-m_N+\frac{g_A}{2}\gamma^\mu\gamma_5
		    u_\mu\right)B\right>, \label{LpiN1}\\
 \mathcal{L}_{\pi
  N}^{(2)}=&c_1\left<\chi_+\right>\left<\bar{B}B\right>-\frac{c_2}{4m_N^2}\left<u_\mu
  u_\nu\right>\left<\bar{B}D^\mu D^\nu N+{\rm h.c.}\right>+\frac{c_3}{2}\left<u_\mu
  u^\mu\right>\left<\bar{B}B\right>\notag\\
&-\frac{c_4}{4}\left<\bar{B}\gamma^\mu\gamma^\nu[u_\mu,u_\nu]
  B\right>+c_5\left<\bar{B}\left(\chi_+-\frac{1}{2}\left<\chi_+\right>\right)B\right>+\left<\bar{B}\sigma^{\mu\nu}\left(\frac{c_6}{2}f_{\mu\nu}^++\frac{c_7}{2}\nu_{\mu\nu}^{(s)}\right)B\right>\label{piN2}.
\end{align}
Here,
\begin{align}
 U&=\exp\left(i\frac{\pi^a\lambda^a}{f}\right),\ \ u=\sqrt{U}=\exp\left(i\frac{\pi^a\lambda^a}{2f}\right)\\
\pi&=\pi^a\lambda^a=
\begin{pmatrix}
 \pi_3+\frac{\eta_8}{\sqrt{3}}&\sqrt{2}\pi^+&\sqrt{2}K^+\\
 \sqrt{2}\pi^-&-\pi_3+\frac{\eta_8}{\sqrt{3}} &\sqrt{2}K^0 \\
 \sqrt{2}K^-&\sqrt{2}\bar{K}^0 &-\frac{2}{\sqrt{3}}\eta_8 
\end{pmatrix}\label{mesfie}\\
 B&=
\begin{pmatrix}
 \frac{\Sigma^0}{\sqrt{2}}+\frac{\Lambda}{\sqrt{6}}&\Sigma^-&p\\
 \Sigma^-&-\frac{\Sigma^0}{\sqrt{2}}+\frac{\Lambda}{\sqrt{6}}&n\\
 \Xi^-&\Xi^0&-\frac{2}{\sqrt{6}}\Lambda 
\end{pmatrix},\label{barfie}\\ 
 \chi&=2B_0\mathcal{M},\ \ \mathcal{M}= 
  \begin{pmatrix}
   m_u&&\\
   &m_d & \\
   & &m_s 
  \end{pmatrix},\\
 f^L_{\mu\nu}&=\partial_\mu l_\nu-\partial_\nu l_\mu-i[l_\mu,l_\nu],\ \ f^R_{\mu\nu}= \partial_\mu r_\nu-\partial_\nu r_\mu-i[r_\mu,r_\nu],\\
  D_\mu&=\partial_\mu+\Gamma_\mu,\ \ \Gamma_\mu=\frac{1}{2}(u^\dagger \partial_\mu u+u\partial_\mu u^\dagger), \label{cov}\\
 u_\mu&= i(u^\dagger\partial_\mu u-u\partial_\mu u^\dagger),\ \ \chi_\pm= u^\dagger\chi u^\dagger\pm u\chi^\dagger u,\\
 f_{\mu\nu}^\pm&= uf^L_{\mu\nu}u^\dagger\pm u^\dagger f^R_{\mu\nu}u,\ \ \nu^{(s)}_{\mu\nu}=\partial_\mu \nu_\nu^{(s)}\pm\partial_\nu \nu_\mu^{(s)},  
\end{align}
and $\left<\cdots\right>$ means the trace in the flavor space.
The Lagrangian that determines the interaction between the hadrons is
constructed so as to be invariant under the chiral transformation of the
hadron fields as
\begin{align}
 U(x)&\mapsto RU(x)L^\dagger,\\
 u(x)&\mapsto \sqrt{RU(x)L^\dagger}\equiv Ru(x)K^{-1}(L,R,U),\ \ K(L,R,U)=\sqrt{RUL^\dagger}R\sqrt{U},\\
 B(x)&\mapsto K(L,R,U)B(x)K^\dagger(L,R,U).
\end{align}
The parameters  $f,\,B_0,\,m_i,\,L_i,\,H_i,\,g_A$, and $c_i$ appearing
in the Lagrangian are
low-energy constants (LECs), the values of  which cannot be fixed
solely from the symmetry and determined  phenomenologically; the values
that are used in our calculation are presented in
Refs.~\cite{Gasser1985b,Bernard1997}.

The relevant degrees of freedom of baryon fields in
Eq.~(\ref{barfie}) are proton and neutron because we are interested in
the medium modification by the nucleon background.
We denote the nucleons in a doublet form as
$N=\,^t(p,n)$.
Although it is known \cite{Hirenzaki2010} (see also, e.g.,
Ref.~\cite{Kelkar2013} for a recent review) that the coupling with
$N^{\ast}$(1535) resonance contributes to the $\eta$ self-energy, the
incorporation of $N^{\ast}$(1535) and other excited baryons with
strangeness is beyond the scope of the present work.
We shall later give a brief comment on possible modification of the
results due to the coupling with $N^{\ast}$(1535).
The meson$-$baryon vertices are derived by expanding $U$ with respect to
the meson fields $\pi^a$.
The explicit forms of the vertices to be used in our calculation are
presented in Appendix \ref{vertices}.

The medium effect is contained in the nucleon propagator $iG(p,k_f)$,
\begin{eqnarray}
 iG(p,k_f)=(\Slash{p}+m_N)\left\{\frac{i}{p^2-m_N^2+i\epsilon}-2\pi\delta(p^2-m_N^2)\delta(p_0)\theta(k_f-|\vec{p}|)\right\},\label{prop}
\end{eqnarray}
where $m_N$ and $k_f$ are the nucleon mass and Fermi momentum,
respectively: The first term of Eq.~(\ref{prop}) is the contribution of
the nucleon propagation in free space and the second term accounts
for the Pauli blocking effect of the nuclear medium.
The number density and the Fermi momentum of the nucleon are related by
$\rho_{p,n}=\frac{k_f^{(p,n)3}}{3\pi^2}$.
The total baryon density $\rho$ and asymmetric density $\delta\rho$ are
defined by $\rho=\rho_n+\rho_p$ and $\delta\rho=\rho_n-\rho_p$,
respectively.
Note that $\delta\rho>0$ means that $\rho_n>\rho_p$ in the
present definition. The nuclear asymmetry is also defined by
$\alpha=\delta \rho/\rho.$

We note that the value of the Fermi momentum $k_f$ around the normal
nuclear density $\rho_0$=0.17fm$^{-3}$ is roughly $2m_\pi$.
In our calculation, we regard $k_f$ as being as small
as the pseudoscalar meson masses and momenta, which are the expansion
parameter in the ordinary chiral perturbation theory.
We call $k_f$ and the masses of the pseudoscalar mesons small
quantities and denote them generically by $q$ in the following.
Hence, our calculation is the expansion with respect to the number of
mesons or nucleon loops, because these loops supply additional
small quantities compared with the tree level.
In addition, we regard the nucleon mass $m_N$ as a large enough
quantity and neglect the ratios of the other quantities to $m_N$.

Here, we note that the the states of the nuclear medium are treated as
a Fermi gas in the leading order in the present formalism, and
accordingly, the nucleon$-$nucleon interaction is switched off
initially.

We calculate the $\eta-3\pi$ decay width up to $O(q^5)$  
in the leading order of the asymmetric density $\delta\rho$.
The final states of the three $\pi$ can be two patterns, i.e.,
$\ppp$ and three $\pi^0$.
The decay amplitude in free space up to $O(q^4)$ is given in
Refs.~\cite{Gasser1985,Bijnens2007}.

\section{The $\eta-\pi^0$ propagator in the asymmetric nuclear
  medium\label{secma}}
This section is devoted to calculation of the $\eta-\pi^0$ propagator
and the $\eta-\pi^0$ mixing angle in the asymmetric nuclear medium.
Here, we denote $\eta$ and $\pi^0$ as the mesons of the mass
eigenstate and $\eta_8$ and $\pi_3$ as the SU(3) eigenstate; the mass
and SU(3) eigenstates are related by Eq.~(\ref{defma}).

The $\eta$ and $\pi^0$ propagator in the asymmetric nuclear medium
$D(p;k_f)$ reads
\begin{align}
 D^{-1}(p;k_f^{(p,n)})=
\begin{pmatrix}
 D_{\eta_8}^{(0)-1}(p)-\Pi_{\eta_8}(k_f^{(p,n)})&-\Pi_{\eta_8\pi_3}(k_f^{(p,n)})\\
 -\Pi_{\eta_8\pi_3}(k_f^{(p,n)})&D_{\pi_3}^{(0)-1}-\Pi_{\pi_3}(k_f^{(p,n)}) 
\end{pmatrix},
\end{align}
where the $D^{(0)}_i(p)$ $(i=\pi_3,\eta_8)$ are the propagators of the
pseudoscalar mesons in the triplet and octet states in free space,
respectively, and the $\Pi_i(k_f)$ is the in-medium self-energy.
The $\Pi_{\eta_8\pi_3}(k_f)$ is the transition amplitude of the $\eta_8$ and
$\pi_3$ mesons.
The meson masses squared $m_{\eta_8}^2$ and $m_{\pi_3}^2$,
are the poles of $D_{\eta_8}$ and $D_{\pi_3}$, respectively, and the
off-diagonal term of the $\eta-\pi^0$ mass matrix $m_{\eta_8\pi_3}^2$
is equal to $\Pi_{\eta_8\pi_3}$.
The $\eta-\pi^0$ mixing angle $\theta$ is obtained in terms of the
masses:
\begin{align}
 \tan2\theta=-\frac{2m_{\eta_8\pi_3}^2}{m_{\eta_8}^2-m_{\pi_3}^2}.\label{madef}
\end{align}
\begin{figure}[t]
 \centering
 \includegraphics[width=10cm]{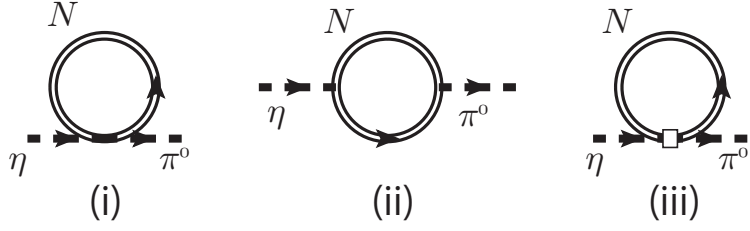}
 \caption{The diagrams contributing to the $\eta$-$\pi^0$ mixing angle in
 the asymmetric nuclear medium.
 The dashed and double-solid lines represent the meson and nucleon
 propagations, respectively.
 The white box at the vertex of diagram (iii) means the
 $\mathcal{L}_{\pi N}^{(2)}$-originated vertex.}
 \label{self_ene1}
\end{figure}

The diagrams that contribute to the $\eta$-$\pi^0$ mixing are shown in
Fig.~\ref{self_ene1}.
We denote the contributions from diagrams (i), (ii), and (iii) as
$\Pi_{\eta_8\pi_3}^{\rm(i)}$, $\Pi_{\eta_8\pi_3}^{\rm(ii)}$, and
$\Pi_{\eta_8\pi_3}^{\rm(iii)}$, respectively.
Actually, $\Pi_{\eta_8\pi_3}^{\rm(i)}$ vanishes because the
$\eta\ppp\bar{N}N$ vertex from $\mathcal{L}_{\pi N}^{(1)}$ is zero, as is
shown in Appendix \ref{secepnn1}.

The $\eta\bar{N}N$ and $\pi^0\bar{N}N$ vertices relevant to
$\Pi_{\eta_8\pi_3}^{\rm(ii)}$ are given in
Eqs.~(\ref{genn}) and (\ref{gpnn}), and we have
\begin{align}
 -i\Pi^{\rm(ii)}_{\eta_8\pi_3}=&-\left(-\frac{g_A}{2\sqrt{3}f}\right)\left(-\frac{g_A}{2f}\right)\int\frac{d^4p}{(2\pi)^4}\tr\{\gamma_5\Slash{k}(\Slash{k}+\Slash{p}+m_N)(-\gamma_5\Slash{k})(\Slash{p}+m_N)\}\notag\\
 &\times
  \left\{\frac{i}{p^2-m_N^2+i\epsilon}-2\pi\delta(p^2-m_N^2)\theta(p_0)\theta(k_f-|\vec{p}|)\right\}\notag\\
 &\times\left\{\frac{i}{(p+k)^2-m_N^2+i\epsilon}-2\pi\delta((p+k)^2-m_N^2)\theta(p_0+k_0)\theta(k_f-|\vec{p}+\vec{k}|)\right\}.
\end{align}
Here, the minus sign on the right-hand side of the first line
comes from the nucleon loop, and the tr in the first line means the
trace of the gamma matrix.
In our calculation, we take the $\eta$-rest frame, so the $\eta$
momentum $k$ is given by $k=(m_\eta,{\bf 0})$.
Eliminating the contribution from the nucleon propagation in
free space, we have
\begin{align}
 -i\Pi_{\eta_8\pi_3}^{\rm(ii)}=&-\frac{1}{\sqrt{3}}\left(\frac{g_A}{2f}\right)^2 4(-2\pi
  i)\int\frac{d^4p}{(2\pi)^4}\left\{-2(k\cdot
			       p)^2+k^2p^2+k^2m_N^2-k^2(k\cdot
			       p)\right\} \notag\\
 &\times\left\{\frac{1}{p^2-m_N^2+i\epsilon}
	   \delta((p+k)^2-m_N^2)\theta(p_0+k_0)\theta(k_f-|\vec{p}+\vec{k}|)\right.\notag\\
&\left.+\frac{1}{(p+k)^2-m_N^2+i\epsilon}\delta(p^2-m_N^2)\theta(p_0)\theta(k_f-|\vec{p}|)\right\}.
\end{align}
Changing the integration variable from $p$ to $p'=p+k$ in the first term
in the brackets, we obtain
\begin{align}
-i\Pi_{\eta_8\pi_3}^{\rm(ii)}=&-\frac{1}{\sqrt{3}}\left(\frac{g_A}{2f}\right)^2 4(-2\pi
  i)\int\frac{d^4p}{(2\pi)^4}\left\{ \frac{-2(k\cdot
			      p)^2+k^2p^2+k^2m_N^2+(k\cdot p)k^2}{(p-k)^2-m_N^2+i\epsilon}\right.\notag\\
&\left.+\frac{-2(k\cdot p)^2+k^2p^2+k^2m_N^2-k^2(k\cdot
   p)}{(p+k)^2-m_N^2+i\epsilon}\right\}\delta(p^2-m_N^2)\theta(p_0)\theta(k_f-|\vec{p}|)\notag\\
 =& -\frac{1}{\sqrt{3}}\left(\frac{g_A}{2f}\right)^2 4(-2\pi
  i)\int\frac{d^3p}{(2\pi)^4}2\times\frac{(k\cdot p)k^2}{-2p\cdot k}\frac{\theta(k_f-|\vec{p}|)}{2E_N(\vec{p})}\notag\\
 =&-i\frac{g_A^2m_\eta^2}{4\sqrt{3}m_N f^2}\rho.
\end{align}
Here, the nucleon mass is treated as a large quantity, and hence
the nucleon energy $E_N(\vec{p})$ is approximated by $m_N$ and the
initial $\eta$ meson is at rest.
Noting that the $\pi^0$ couples to a proton and a neutron with opposite
signs, we obtain the final form as
\begin{eqnarray}
 -i\Pi_{\eta_8\pi_3}^{\rm(ii)}=i\frac{g_A^2m_\eta^2}{4\sqrt{3}m_N
  f^2}\delta\rho.\label{38b}
\end{eqnarray}

Now, we decompose $\Pi_{\eta_8\pi_3}^{\rm(iii)}$ into
$\Pi_{\eta_8\pi_3}^{\rm(iii1)}$ and $\Pi_{\eta_8\pi_3}^{\rm(iii5)}$, which
come from the terms proportional to $c_1$ and $c_5$ in the chiral
Lagrangian in Eq.~(\ref{piN2}), respectively.
The $\eta\ppp\bar{N}N$ vertex proportional to $c_1$ contained in
$\mathcal{L}_{\pi N}^{(2)}$ is given in Eq.~(\ref{gepnnc1}),
and thus we have for $\Pi_{\eta_8\pi_3}^{\rm(iii1)}$
\begin{align}
 -i\Pi_{\eta_8\pi_3}^{\rm(iii1)}=&-\left(i\frac{4c_1m_1^2}{\sqrt{3}f^2}\right)\int\frac{d^4p}{(2\pi)^4}\tr(\Slash{p}+m_N)\left\{\frac{i}{p^2-m_N^2+i\epsilon}-2\pi\delta(p^2-m_N^2)\theta(p_0)\theta(k_f-|\vec{p}|))\right\},\notag
\end{align}
which is reduced to
\begin{eqnarray}
 -i\Pi_{\eta_8\pi_3}^{\rm(iii1)}=i\frac{4c_1m_1^2}{\sqrt{3}f^2}\rho.\label{38c1}
\end{eqnarray}

With the use of the $\eta\pi^0\bar{N}N$ vertex coming from the $c_5$ term
in Eq.~(\ref{gepnnc5}), $\Pi_{\eta_8\pi_3}^{\rm(iii5)}$ is reduced to
\begin{align}
 -i\Pi^{\rm(iii5)}_{\eta_8\pi_3}=&-\left(-i\frac{4c_5B_0m_u}{\sqrt{3}f^2}\right)\int\frac{d^4p}{(2\pi)^4}\tr(\Slash{p}+m_N)\notag\\
&\hspace{60pt}\times\left[\frac{i}{p^2-m_N^2+i\epsilon}-2\pi\delta(p^2-m_N^2)\theta(p_0)\theta(k_f^{(p)}-|\vec{p}|)\right]\notag\\
 &-\left(i\frac{4c_5B_0m_d}{\sqrt{3}f^2}\right)\int\frac{d^4p}{(2\pi)^4}\tr(\Slash{p}+m_N)\notag\\
&\hspace{60pt}\times\left[\frac{i}{p^2-m_N^2+i\epsilon}-2\pi\delta(p^2-m_N^2)\theta(p_0)\theta(k_f^{(n)}-|\vec{p}|)\right]\notag\\
 &-\frac{1}{2}\left(-i\frac{4c_5m_1^2}{\sqrt{3}f^2}\right)\int\frac{d^4p}{(2\pi)^4}\tr(\Slash{p}+m_N)\notag\\
&\hspace{60pt}\times\left\{\left[\frac{i}{p^2-m_N^2+i\epsilon}-2\pi\delta(p^2-m_N^2)\theta(p_0)\theta(k_f^{(p)}-|\vec{p}|)\right]\right.\notag\\
 &\hspace{70pt}\left.+ \left[\frac{i}{p^2-m_N^2+i\epsilon}-2\pi\delta(p^2-m_N^2)\theta(p_0)\theta(k_f^{(n)}-|\vec{p}|)\right]\right\},
\end{align}
which tells us that the contribution from the nuclear medium is given by
\begin{align}
 -i\Pi_{\eta_8\pi_3}^{\rm(iii5)}=&i\frac{4c_5}{2\sqrt{3}f^2}B_0(m_u+m_d)\delta\rho\notag\\
 =&i\frac{2c_5m_\pi^2}{\sqrt{3}f^2}\delta\rho. \label{38c5}
\end{align}

Incorporating the contribution in free space and the nuclear medium
effect given in Eqs.~(\ref{38b}), (\ref{38c1}), and (\ref{38c5}),
we have, for the in-medium $\eta$-$\pi^0$ mixing angle
$\theta^{(\rho)}$,
\begin{align}
  \tan2\theta^{(\rho)}=\frac{2}{m_\eta^2-m_{\pi^0}^2}\left(\frac{m_1^2}{\sqrt{3}}+\left(\frac{g_A^2m_\eta^2}{4\sqrt{3}f^2m_N}+\frac{2c_5m_{\pi}^2}{\sqrt{3}f^2}\right)\delta\rho+\frac{4c_1m_1^2}{\sqrt{3}f^2}\rho\right),\label{maeq}
\end{align}
where 
\begin{align}
 m_1^2=B_0(m_d-m_u)=m_{K^0}^2-m_{K^+}^2-m_{\pi^0}^2+m_{\pi^+}^2.
\end{align}
Equation (\ref{maeq}) shows that the in-medium mixing angle depends not
only on the asymmetric density $\delta\rho$ appearing in the second
term, but also on the total baryon density $\rho$ in the third term.
We see that $\delta\rho$ enhances the mixing angle in the neutron-rich
asymmetric nuclear medium, while $\rho$ reduces
the mixing angle because the coefficient $c_1$ is negative.
The parameter $c_1$ is determined so as to reproduce
the low-energy $\pi N$ scattering \cite{Bernard1997} and the sign
reflects the nature of the low-energy $\pi N$ interaction.
The resultant decay width is obtained from the balance of these
effects.
From the assignment of the isospin, the proton and
neutron densities affect the decay
in the same way as the $u$ and $d$ quark mass do.
A larger density difference of the $u$ and $d$ quarks means a
stronger violation of the isospin symmetry, so the $\eta$-$\pi^0$ mixing
angle is enhanced in the neutron-rich nuclear medium.
\begin{figure}[t]
\begin{minipage}[t]{0.475\hsize}
\centering
  \includegraphics[width=6cm]{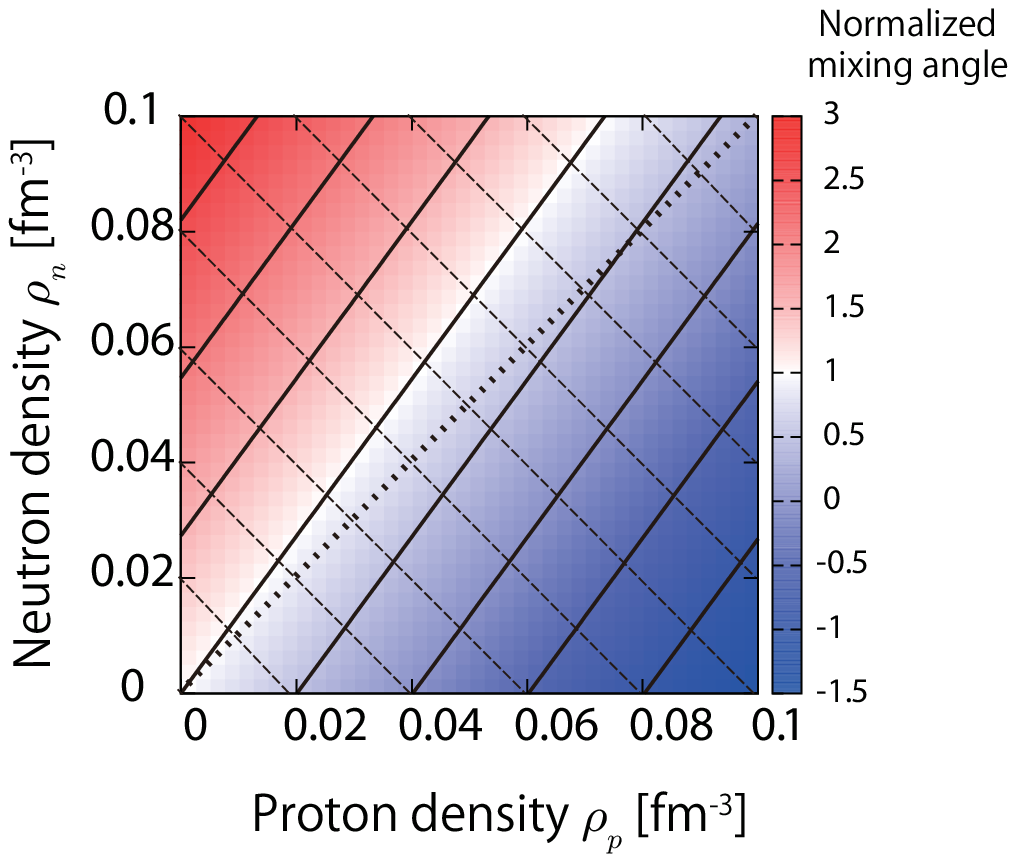}
  \caption{The $\eta$-$\pi^0$ mixing angle in the asymmetric nuclear
 medium up to $O(q^5)$. 
 The solid lines are the contour lines and are plotted per 0.5 of the
 mixing angle normalized with the free-space value.
 The horizontal and vertical axes are the proton density,
 $\rho_p$[fm$^{-3}$], and the neutron density, $\rho_n$[fm$^{-3}$].
 The dashed and dotted lines are the constant-$\rho$ and
 vanishing-$\delta\rho$ lines, respectively.
 The lower-left and upper-right regions of the figure correspond to
 small and large total baryon densities $\rho$, and the upper and lower
 sides  of the dotted line are the neutron- and proton-rich regions,
 respectively.
 The value is normalized by the mixing angle in free space $\theta^{(0)}
 \simeq 1.058\times 10^{-2}$\, [rad].
 The mixing angle is smaller than that of the free-space value in the blue
 region where $\rho_p$ is large, and larger in the red region where
 $\rho_n$ is large.}
 \label{ma}
\end{minipage}
\hfill
\begin{minipage}[t]{0.475\hsize}
  \centering
 \includegraphics[width=6cm]{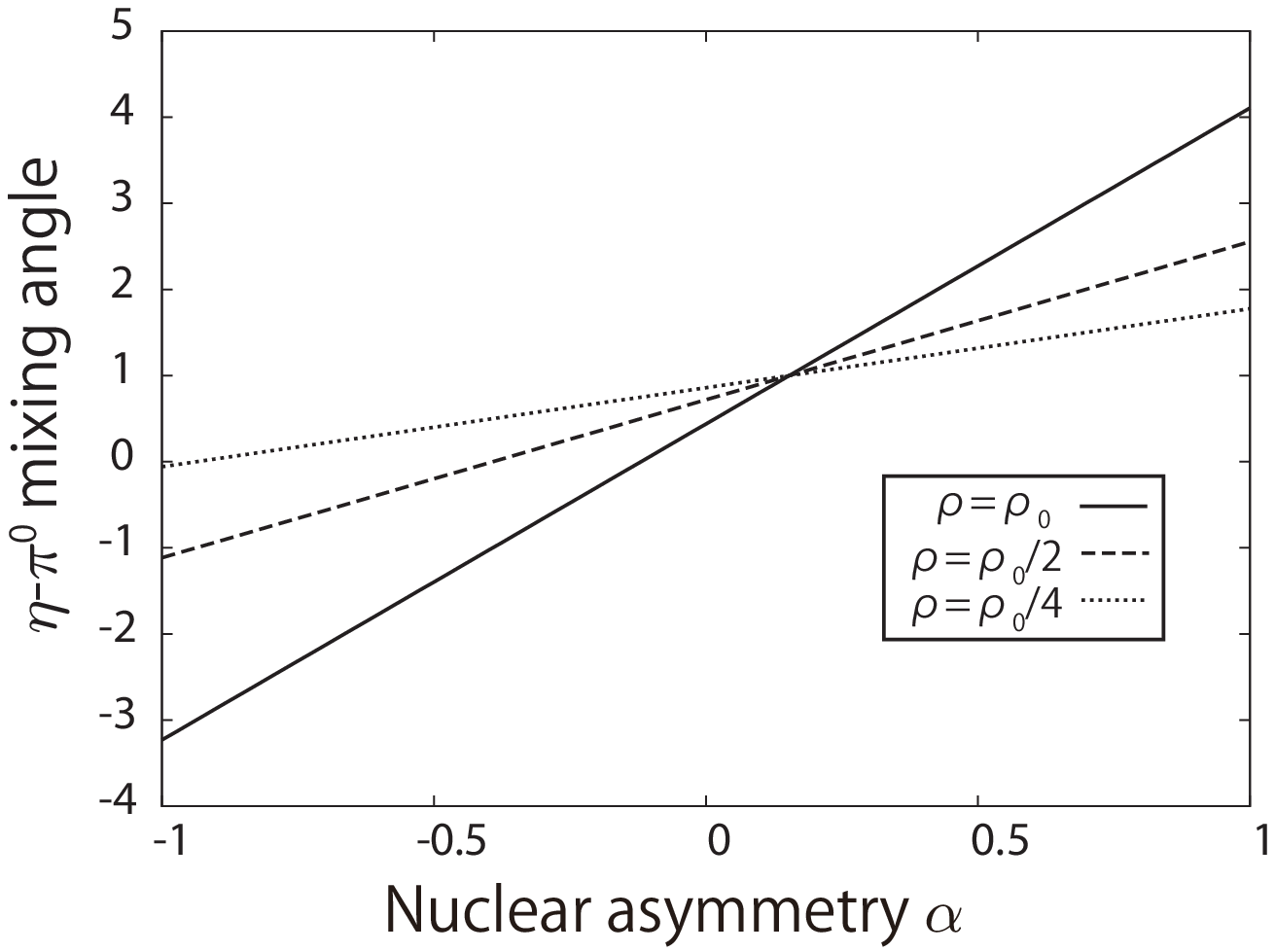}
 \caption{The nuclear asymmetry $\alpha$ dependence of the
 $\eta-\pi^0$ mixing angle in the nuclear medium.
 The horizontal and vertical axes represent the nuclear asymmetry and
 the $\eta-\pi^0$ mixing angle normalized by the value at
 $\rho=\delta\rho=0$.
 The solid, dashed, and dotted lines are the lines with
 $\rho=\rho_0$, $\rho_0/2$, and $\rho_0/4$, respectively.}
 \label{ma_alpha}
\end{minipage}
\end{figure}

A contour plot of the $\eta-\pi^0$ mixing angle in the asymmetric nuclear
medium is presented in Fig.~\ref{ma}. In the present work,  we  use the
following values in the numerical calculation \cite{Bernard1997}:
$c_1=-0.93\pm 0.10$ GeV$^{-1}$, $c_5=-0.09\pm 0.01$ GeV$^{-1}$, and
$f=93$ MeV.
We note that the $c_i$ of the LECs have uncertainties of some 10\%.
The masses of all the hadrons are taken to be the experimental values
listed in Ref.~\cite{Beringer2012} and thus $m_1^2=5165.86$ MeV$^2.$  
One can find from this figure that the
$\eta-\pi^0$ mixing angle is enhanced in the neutron-rich asymmetric
nuclear medium, and tends to be slightly suppressed by the total baryon
density.

We show the nuclear asymmetry $\alpha$ dependence of the
$\eta-\pi^0$ mixing angle in Fig.~\ref{ma_alpha}:
One sees that the mixing angle is enhanced by $\alpha$, and the
slope of the $\alpha$ dependence is bigger for the higher total baryon
density.
In fact, the slope of the mixing angle in terms of $\alpha$ in the small
density is given as
\begin{align}
 \frac{d\theta}{d\alpha}\sim\frac{1}{2}\frac{d\tan 2\theta}{d\alpha}=\frac{\rho/\sqrt{3}f^2}{m_\eta^2-m_{\pi^0}^2}\left\{-4c_1m_1^2\alpha+\left(\frac{g_A^2m_\eta^2}{4m_N}+2c_5m_\pi^2\right)\right\},
\end{align}
which is proportional to $\rho$.

The density dependence of the $\eta-\pi^0$ mixing angle has an
uncertainty coming from those of the LECs, $c_1$ and $c_5$
\cite{Bernard1997}.
The resultant uncertainty of the mixing angle is about 10\%.

\section{The $\eta-3\pi$ decay width in the
 asymmetric nuclear medium\label{sece3p}}
In this section, we estimate the $\eta-3\pi$ decay width in the
asymmetric nuclear medium.
The partial width $\Gamma$ in the rest frame of a particle with mass $M$
reads 
\begin{eqnarray}
\Gamma=\frac{1}{n!}\frac{1}{4M}\int ds\int dt
 |\mathcal{M}|^2.\label{widthdef}
\end{eqnarray}
Here, $n$ is the number of identical particles in the final state,
$\mathcal{M}$ the matrix element of the decay, and
$s=(p_\eta-p_{\pi^0})^2$, $t=(p_\eta-p_{\pi^+})^2$ the Mandelstam
 variables.
\begin{figure}[t]
 \centerline{\includegraphics[width=14cm]{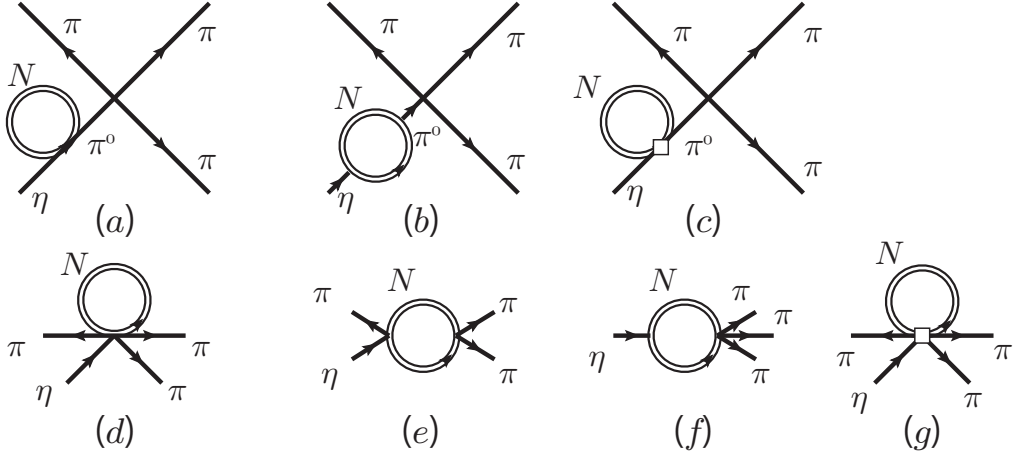}}
\caption{The diagrams contributing to the $\eta$ decay into three
 $\pi$.
 The meanings of the lines and vertices are same as
 in Fig.~\ref{self_ene1}.
 Diagrams ($a$), ($b$), and ($c$) are the contributions from the mixing
 angle and ($d$) to ($g$) give the medium effects on the $\eta-3\pi$
 decay amplitude directly.}
\label{diagrams}
\end{figure}

The diagrams contributing to the $\eta-3\pi$ decay amplitude
are shown in Fig.~\ref{diagrams}.
Diagrams ($a$), ($b$), and ($c$) affect the decay amplitude
through the $\eta-\pi^0$ mixing angle, and
diagrams ($d$) to ($g$) give the medium effects on the decay amplitude
directly.
The in-medium $\eta-\pi^0$ mixing angle has already been calculated in
Sect.~\ref{secma}, i.e., the contributions from diagrams ($a$), ($b$),
and ($c$).
We evaluate diagrams ($d$) to ($g$) in Fig.~\ref{diagrams} and the
decay width in the asymmetric nuclear medium in this section.
In Sect.\ref{secppp} and \ref{sec3p}, we calculate the decay amplitude of
$\eta$ into $\ppp$ and 3$\pi^0$, respectively, and we show the results
of the numerical estimation of the decay width in Sect.~\ref{secnum}.

\subsection{The $\eta$ decay into $\ppp$\label{secppp}}
In this subsection, we calculate the $\eta$ decay amplitude into $\ppp$
in the asymmetric nuclear medium.
We show the matrix elements of $\eta$ decay into $\ppp$ that come from
the diagrams ($d$) to ($g$) in Fig.~\ref{diagrams}.

Diagrams ($d$) and ($e$) do not contribute to the decay
amplitude because both the $\eta\pi^0\pi^+\pi^-\bar{N}N$ and
the $\eta\pi^0\bar{N}N$ vertices coming from
$\mathcal{L}_{\pi N}^{(1)}$ are equal to zero, as shown in
Eq.~(\ref{gepppnn1}) and Appendix \ref{secepnn1}.

With the use of the vertices of $\eta\bar{N}N$ and $\ppp\bar{N}N$ given in
Eqs.~(\ref{genn}) and (\ref{gpppnn}), the contribution from diagram
($f$) is written as 
 \begin{align}
 i\mathcal{M}^{(f)}=&(-)\left(-\frac{g_A}{2\sqrt{3}f}\right)\left(-\frac{g_A}{24f^3}\right)\int\frac{d^4p}{(2\pi)^4}\tr\{\gamma_5\Slash{k}(\Slash{p}+\Slash{k}+m_N)\gamma_5(2\Slash{p}_0-\Slash{p}_+-\Slash{p}_-)(\Slash{p}+m_N)\}\notag\\
 &\times\left\{\frac{i}{p^2-m_N^2+i\epsilon}-2\pi\delta(p^2-m_N^2)\theta(p_0)\theta(k_f-|\vec{p}|)\right\}\notag\\
& \times \left\{\frac{i}{(p+k)^2-m_N^2+i\epsilon}-2\pi\delta((p+k)^2-m_N^2)\theta(p_0+k_0)\theta(k_f-|\vec{p}+\vec{k}|)\right\}.
\end{align}
The minus sign comes from the fermion loop.
Eliminating the pure-vacuum contributions and setting $A=2p_0-p_+-p_-$, we have
\begin{align}
  i\mathcal{M}^{(f)}=&\frac{ig_A^2}{48\sqrt{3}f^4(2\pi)^3}\int
  d^4p 4\{2(k\cdot p)(A\cdot p)+k^2(A\cdot p)-p^2(A\cdot k)-m_N^2(A\cdot k) \}\\
&\times \left\{\frac{1}{p^2-m^2_N}\theta(p_0+k_0)\delta((p+k)^2-m_N^2)\theta(k_f-|\vec{p}+\vec{k}|)\right.\notag\\
   &\left.+\frac{1}{(p+k)^2-m_N^2}\theta(p_0)\delta(p^2-m_N^2)\theta
   (k_f-|\vec{p}|)\right\}.
\end{align}
Changing the integration variable of the first term in the bracket $p$
into $p'=p+k$ and keeping the leading order of the momentum,
$i\mathcal{M}^{(f)}$ is reduced to
\begin{eqnarray}
i\mathcal{M}^{(f)}=\frac{ig_A^2}{48\sqrt{3}f^4(2\pi)^3}\int
 d^3 p\frac{2k^2(A\cdot k)}{2p\cdot k}\frac{1}{2E_N(\vec{p})}\times \theta(k_f-|\vec{p}|).
\end{eqnarray}
Approximating $E_N(\vec{p})$ to $m_N$, we arrive at
\begin{align}
  i\mathcal{M}^{(f)}=i\frac{g_A^2}{48\sqrt{3}f^4}A_0\rho .
\end{align}
Using the energy conservation, $A_0=3E_{\pi^0}-m_\eta$, and taking
account of the opposite sign of the vertex between $\pi$ and the proton
or neutron, we find that the leading contribution containing the density
effect finally has the following form:
\begin{eqnarray}
 i\mathcal{M}^{(f)} =&i\frac{g_A^2}{48\sqrt{3}f^4}m_\eta\delta\rho-i\frac{g_A^2}{16\sqrt{3}f^4}E_{\pi^0}\delta\rho.\label{diagf}
\end{eqnarray}

Here, we decompose the contribution from diagram ($g$) in
Fig.~\ref{diagrams} in two parts; the term proportional to $c_1$ and
$c_5$, respectively:
$\mathcal{M}^{(g)}$ as
$\mathcal{M}^{(g)}=\mathcal{M}^{(g1)}+\mathcal{M}^{(g5)}$, where
$\mathcal{M}^{(g1)}$ and $\mathcal{M}^{(g5)}$ are proportional to $c_1$
and $c_5$.
The $\eta\ppp\bar{N}N$ vertices proportional to $c_1$ and $c_5$ are
presented in Eqs.~(\ref{gepppnnc1}) and (\ref{gm3p0nn}), respectively.

With the $\eta\ppp\bar{N}N$ vertex given in Eq.~(\ref{gepppnnc1}), 
$\mathcal{M}^{(g1)}$ is given as
\begin{align}
 i\mathcal{M}^{(g1)}=&-\left(-i\frac{4c_1m_1^2}{3\sqrt{3}f^4}\right)
\int\frac{d^4p}{(2\pi)^4}\tr(\Slash{p}+m_N)\notag\\
&\hspace{80pt}\times\left\{\frac{i}{p^2-m_N^2+i\epsilon}
-2\pi\delta(p^2-m_N^2)\theta(p_0)\theta(k_f-|\vec{p}|)\right\}.
\end{align}
Eliminating the vacuum part of the nucleon propagator,
$\mathcal{M}^{(g1)}$ is reduced to
\begin{eqnarray}
 i\mathcal{M}^{(g1)}=-i\frac{4c_1m_1^2}{3\sqrt{3}f^4}\rho.
\end{eqnarray}

Using the $\eta3\pi^0\bar{N}N$ vertex given in Eq.~(\ref{gm3p0nn}),
$\mathcal{M}^{(g5)}$ is written as
\begin{align}
i\mathcal{M}^{(g5)}=&-\left(i\frac{4c_5m_u}{3\sqrt{3}}\right)\int\frac{d^4p}{(2\pi)^4}\tr(\Slash{p}+m_N)\notag\\
&\hspace{80pt}\times\left\{\frac{i}{p^2-m_N^2+i\epsilon}-2\pi\delta(p^2-m_N^2)\theta(p_0)\theta(k_f^{(p)}-|\vec{p}|)\right\}\notag\\
&-\left(-i\frac{4c_5m_d}{3\sqrt{3}}\right)\int\frac{d^4p}{(2\pi)^4}\tr(\Slash{p}+m_N)\notag\\
&\hspace{80pt}\times\left\{\frac{i}{p^2-m_N^2+i\epsilon}-2\pi\delta(p^2-m_N^2)\theta(p_0)\theta(k_f^{(n)}-|\vec{p}|)\right\}\notag\\
&-\frac{1}{2}\left(i\frac{4c_5m_1^2}{3\sqrt{3}f^4}\right)\int\frac{d^4p}{(2\pi)^4}\tr(\Slash{p}+m_N)\notag\\
&\hspace{80pt}\times\left[\left\{\frac{i}{p^2-m_N^2+i\epsilon}-2\pi\theta(p_0)\delta(p^2-m_N^2)\theta(k_f^{(p)}-|\vec{p}|)\right\}\right.\notag\\
&\hspace{90pt}\left.+\left\{\frac{i}{p^2-m_N^2+i\epsilon}-2\pi\theta(p_0)\delta(p^2-m_N^2)\theta(k_f^{(n)}-|\vec{p}|)\right\}\right].
\end{align}
Omitting the terms that come from the free-space propagation of the nucleon,
we obtain $\mathcal{M}^{(g5)}$ as
\begin{align}
 i\mathcal{M}^{(g5)}=-i\frac{2c_5m_\pi^2}{3\sqrt{3}f^4}\delta\rho.
\end{align}
Thus, $\mathcal{M}^{(g)}$ is written as
\begin{align}
 i\mathcal{M}^{(g)}=-i\frac{4c_1m_1^2}{3\sqrt{3}f^4}\rho-i\frac{2c_5m_\pi^2}{3\sqrt{3}f^4}\delta\rho.\label{diagg}
\end{align}

Taking account of the contributions from free space and the
modification of the $\eta-\pi^0$ mixing angle,
we have the matrix element of the $\eta$ decay into $\ppp$ in
the asymmetric nuclear medium up to $O(q^5)$ as
\begin{align}
& \mathcal{M}_{\eta\rightarrow\pi^0\pi^+\pi^-}=
-\left(\frac{m_1^2}{3\sqrt{3}f^2}+\frac{s-s_0}{f^2}\sin\theta^{(\rho)}\right)
+\mathcal{M}_{\eta\rightarrow\pi^0\pi+\pi^-}^{(4)}\notag\\
&\ \ \ \ =-\frac{m_1^2}{3\sqrt{3}f^2}\left(1+\frac{3(s-s_0)}{m_\eta^2-m_{\pi^0}^2}\right)
+\sin\theta^{(0)}\mathcal{M}^{\rm(4)vac}_{\eta\rightarrow\pi^0\pi^+\pi^-} \notag\\
&\ \ \ \ \ \ +\left\{-\frac{s-s_0}{f^2}\frac{1}{m_\eta^2-m_{\pi^0}^2}
\left(\frac{g_A^2m_\eta^2}{4\sqrt{3}f^2}+\frac{2c_5m_\pi^2}{\sqrt{3}f^2}\right)
+\frac{g_A^2}{48\sqrt{3}f^4}(m_\eta-E_{\pi^0})-\frac{2c_5m_\pi^2}{3\sqrt{3}f^4}\right\}\delta\rho \notag\\
&\ \ \ \ \ \ -\frac{4c_1\rho}{f^2}\frac{m_1^2}{3\sqrt{3}f^2}\left(1+\frac{3(s-s_0)}{m_\eta^2-m_{\pi^0}^{2}}\right), \label{mppp2}
\end{align}
where
\begin{align}
 \mathcal{M}^{(4)}_{\eta\rightarrow\pi^0\pi^+\pi^-}
 =&\sin\theta^{(0)}\mathcal{M}^{\rm(4)vac}_{\eta\rightarrow
\ppp}+\mathcal{M}^{\rm(f)}+\mathcal{M}^{(g)}.
\end{align}
Here, $s_0$ is given as $s_0=m_\eta^2/3+m_\pi^2$, and
$\mathcal{M}^{\rm(4)vac}_{\eta\rightarrow\ppp}$ is the meson one-loop
contribution in free space, which is known to give a large
contribution, as mentioned in Sect.~\ref{intro}.
The details of the form and calculation of
$\mathcal{M}^{\rm(4)vac}_{\eta\rightarrow\ppp}$ are given in
Refs.~\cite{Gasser1985,Bijnens2007}.
$\mathcal{M}^{\rm(f)}$ and $\mathcal{M}^{(g)}$ are
given in Eqs.~(\ref{diagf}) and (\ref{diagg}), respectively.
We denote the $\eta-\pi^0$ mixing angle in free space by
$\theta^{(0)}$, which is given by setting $\rho=\delta\rho=0$ in
Eq.~(\ref{maeq}).
In this calculation, we have assumed that the isospin symmetry
breaking is so small that we can make the approximation that
$\sin\theta\sim\theta\sim\tan2\theta/2$.

\subsection{The $\eta$ decay into three $\pi^0$\label{sec3p}}
In this subsection, we give the decay amplitude for the $\eta$ decay to
three $\pi^0$ in the asymmetric nuclear medium.

First of all, the contributions from diagrams ($d$), ($e$), and ($f$)
vanish in the three-$\pi^0$ case.
Diagrams ($d$) and ($e$) give no contribution in the same way as in
the case of the $\ppp$ decay.
The contribution of diagram ($f$) also vanishes because the
$3\pi^0\bar{N}N$ vertex is zero, as is shown in
Appendix~\ref{secmmmnn}.
Thus, the decay amplitude in the asymmetric nuclear medium is solely
given by the sum of the contributions from the diagram ($g$).
$\mathcal{M}^{(g1)}$, which is proportional to $c_1$, is written as
\begin{align}
 i\mathcal{M}^{(g1)}=-\left(-i\frac{2c_1m_1^2}{3\sqrt{3}f^4}\right)\int\frac{d^4p}{(2\pi)^4}\left\{\frac{i}{p^2-m_N^2+i\epsilon}-2\pi\delta(p^2-m_N^2)\theta(p_0)\theta(k_f-|\vec{p}|)\right\},
\end{align}
where the $\eta3\pi^0\bar{N}N$ vertex appearing from $c_1$ is presented in
Eq.~(\ref{ge3p0nnc1}).
The nuclear medium modification is evaluated to be
\begin{align}
 i\mathcal{M}^{(g1)}=&-i\frac{2m_1^2c_1}{3\sqrt{3}f^4}\rho.\label{mg1}
\end{align}

The $\eta3\pi^0\bar{N}N$ vertex with $c_5$ is given in
Eq.~(\ref{gm3p0nn}), and the term proportional to $c_5$ in
$\mathcal{M}^{(g5)}$ reads
\begin{align}
 i\mathcal{M}^{(g5)}=&(-)i\frac{2c_5B_0}{3\sqrt{3}f^4}\int\frac{d^4p}{(2\pi)^4}\tr(\Slash{p}+m_N)\notag\\
&\hspace{70pt}\times\left(m_u\left\{\frac{i}{p^2-m_N^2+i\epsilon}-2\pi\delta(p^2-m_N^2)\theta(p_0)\theta(k_f^{(p)}-|\vec{p}|)\right\}\right.\notag\\
&\hspace{80pt}\left.-m_d\left\{\frac{i}{p^2-m_N^2+i\epsilon}-2\pi\delta(p^2-m_N^2)\theta(p_0)\theta(k_f^{(n)}-|\vec{p}|)\right\}\right)\notag\\
&-\frac{1}{2}\left(i\frac{2c_5m_1^2}{3\sqrt{3}f^4}\right)\int\frac{d^4p}{(2\pi)^4}\tr(\Slash{p}+m_N)\notag\\
&\hspace{70pt}\times\left(\left\{\frac{i}{p^2-m_N^2+i\epsilon}-2\pi\delta(p^2-m_N^2)\theta(p_0)\theta(k_f^{(p)}-|\vec{p}|)\right\}\right.\notag\\
&\hspace{80pt}\left.+\left\{\frac{i}{p^2-m_N^2+i\epsilon}-2\pi\delta(p^2-m_N^2)\theta(p_0)\theta(k_f^{(n)}-|\vec{p}|)\right\}\right).
\end{align}
Omitting the free-space part in the right-hand side of the equation, we
obtain
\begin{align}
  i\mathcal{M}^{(g5)}=&-i\frac{c_5m_{\pi^0}^2}{3\sqrt{3}f^4}\delta\rho.\label{mg5}
\end{align}
Summing up Eqs.~(\ref{mg1}) and (\ref{mg5}), we have
\begin{align}
 i\mathcal{M}^{(g)}=-i\frac{2m_1^2c_1}{3\sqrt{3}}\rho-i\frac{c_5m_{\pi}^2}{3\sqrt{3}f^4}\delta\rho.
\end{align}

Thus we have the matrix element of the $\eta$ decay into $\p$ as
\begin{align}
\mathcal{M}_{\eta\rightarrow3\pi^0}=&-\frac{m_1^2}{\sqrt{3}f^2}+\mathcal{M}^{(4)}_{\eta\rightarrow
 3\pi^0} \nonumber \\
 =&-\frac{m_1^2}{\sqrt{3}f^2}+\sin\theta^{(0)}\mathcal{M}^{\rm (4)vac}_{\eta\rightarrow 3\pi^0}-\frac{2m_1^2c_1}{3\sqrt{3}f^4}\rho-\frac{c_5m_\pi^2}{3\sqrt{3}f^4}\delta\rho,
\label{m3pi0}
\end{align}
with
\begin{align}
 \mathcal{M}^{(4)}_{\eta\rightarrow3\pi^0}
 =&\sin\theta^{(0)}\mathcal{M}^{\rm(4)vac}_{\eta\rightarrow3\pi^0}+\mathcal{M}^{(g)}.
\end{align}
Here, $\mathcal{M}^{\rm(4)vac}_{\eta\rightarrow3\pi^0}$ is the
contribution from the meson one-loop and its detailed form is presented in
Refs.~\cite{Gasser1985,Bijnens2007}.
We note that the amplitude does not depend on the $\eta-\pi^0$
mixing angle in the leading order, on account of the symmetry of the
final state consisting of three identical $\pi^0$.
For this reason, the medium modification of the $\eta$ decay
into three $\pi^0$ is small, which will be demonstrated in the
numerical calculation given in the next subsection.

\subsection{Numerical results\label{secnum}}
Using the definition of the decay width given in
Eq.~(\ref{widthdef}) and the matrix elements presented in Eqs.~(\ref{mppp2})
and (\ref{m3pi0}), we evaluate the partial width of the $\eta$ decay
into 3$\pi$.

\begin{figure}[t]
\begin{minipage}[t]{0.475\hsize}
 \centering
 \includegraphics[width=7.8cm]{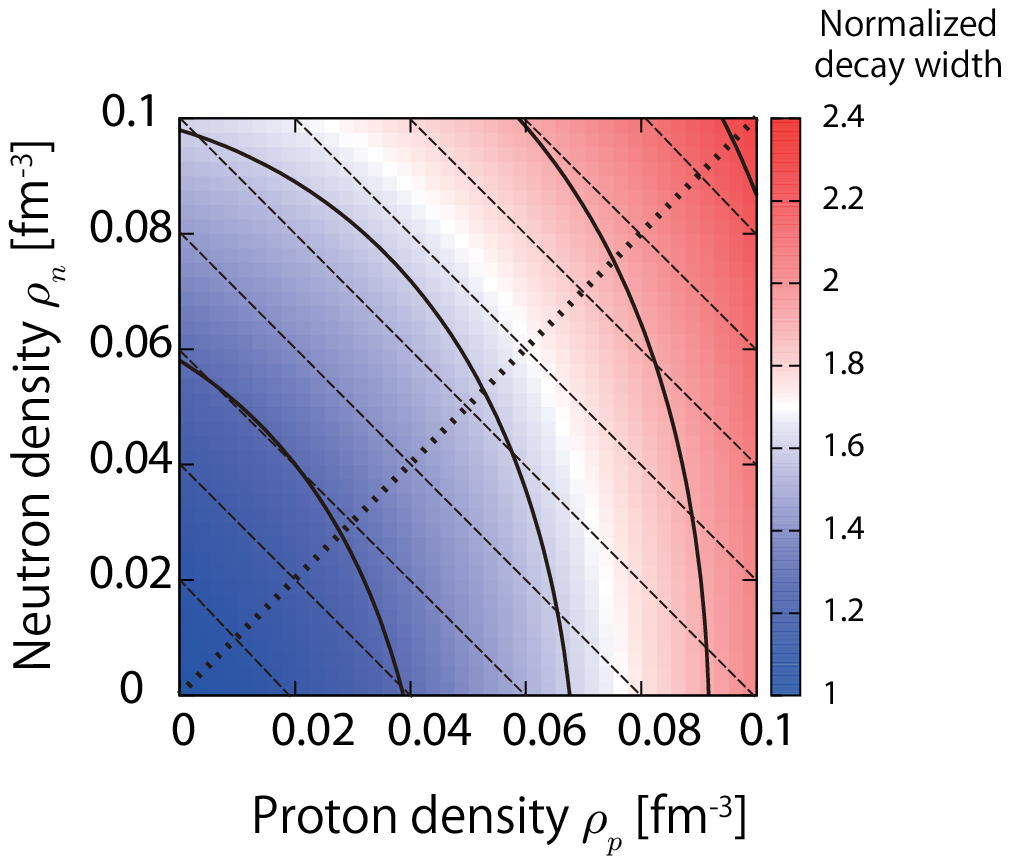}
 \caption{$\eta\rightarrow\pi^0\pi^+\pi^-$ decay width in asymmetric
 nuclear medium up to $O(q^5)$.
 The horizontal and vertical axes represent $\rho_p$ and
 $\rho_n$, respectively.
 The solid line represents the contour of the decay width.
 The dashed and dotted lines mean the constant-$\rho$ and
 the vanishing-$\delta\rho$ lines, respectively.
 The width is normalized
 by the value at $\rho=\delta\rho=0$.
 The contour is plotted per 0.3.
 The width in free space is 163 eV.}
 \label{widthppp}
\end{minipage}
\hfill
\begin{minipage}[t]{0.475\hsize}
  \centering
 \includegraphics[width=6cm]{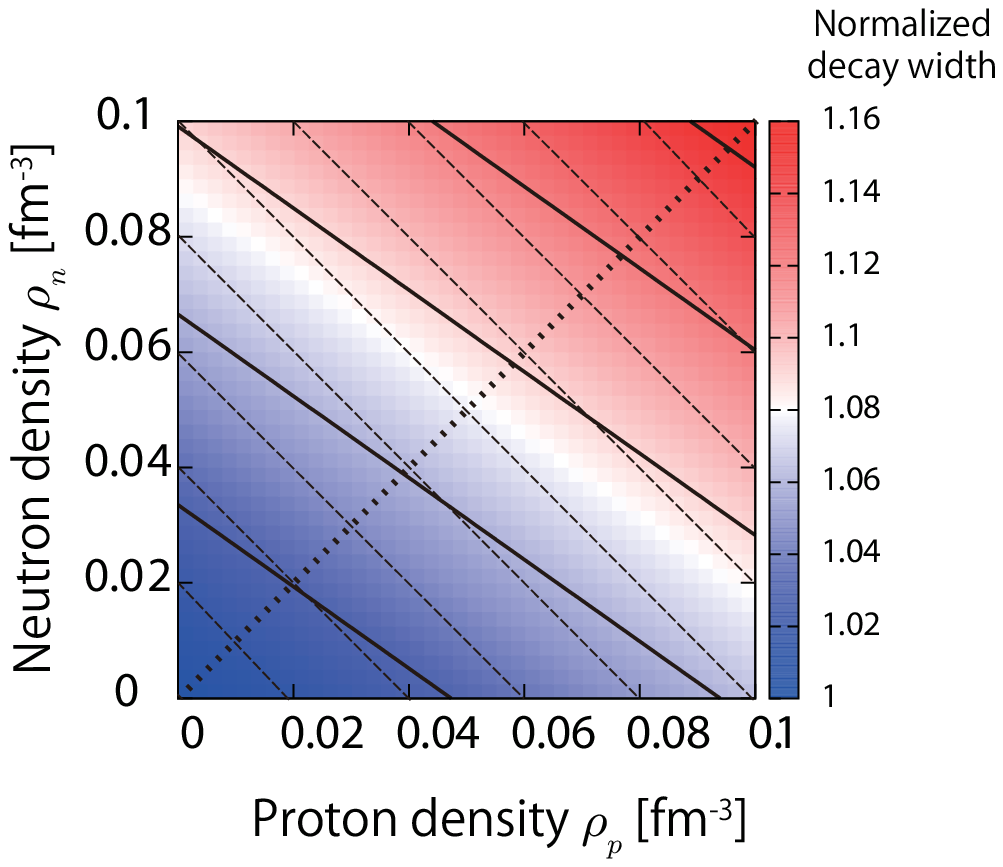}
 \caption{$\eta\rightarrow 3\pi^0$ decay width in the
 asymmetric nuclear medium up to $O(q^5)$.
 The axes and the lines are the same as those in Fig.~\ref{widthppp}.
 The contour is plotted per 0.03.
 The width in free space is 298 eV.}
 \label{width3pi0}
\end{minipage}
\end{figure}

Figures \ref{widthppp} and \ref{width3pi0} are contour plots of the
decay width of $\eta$ into $\ppp$ and into three $\pi^0$, respectively.
The widths are normalized with each value at $\rho=\delta\rho=0$,
respectively.

\begin{figure}[t]
\begin{minipage}[t]{0.475\hsize}
 \centering
 \includegraphics[width=6cm]{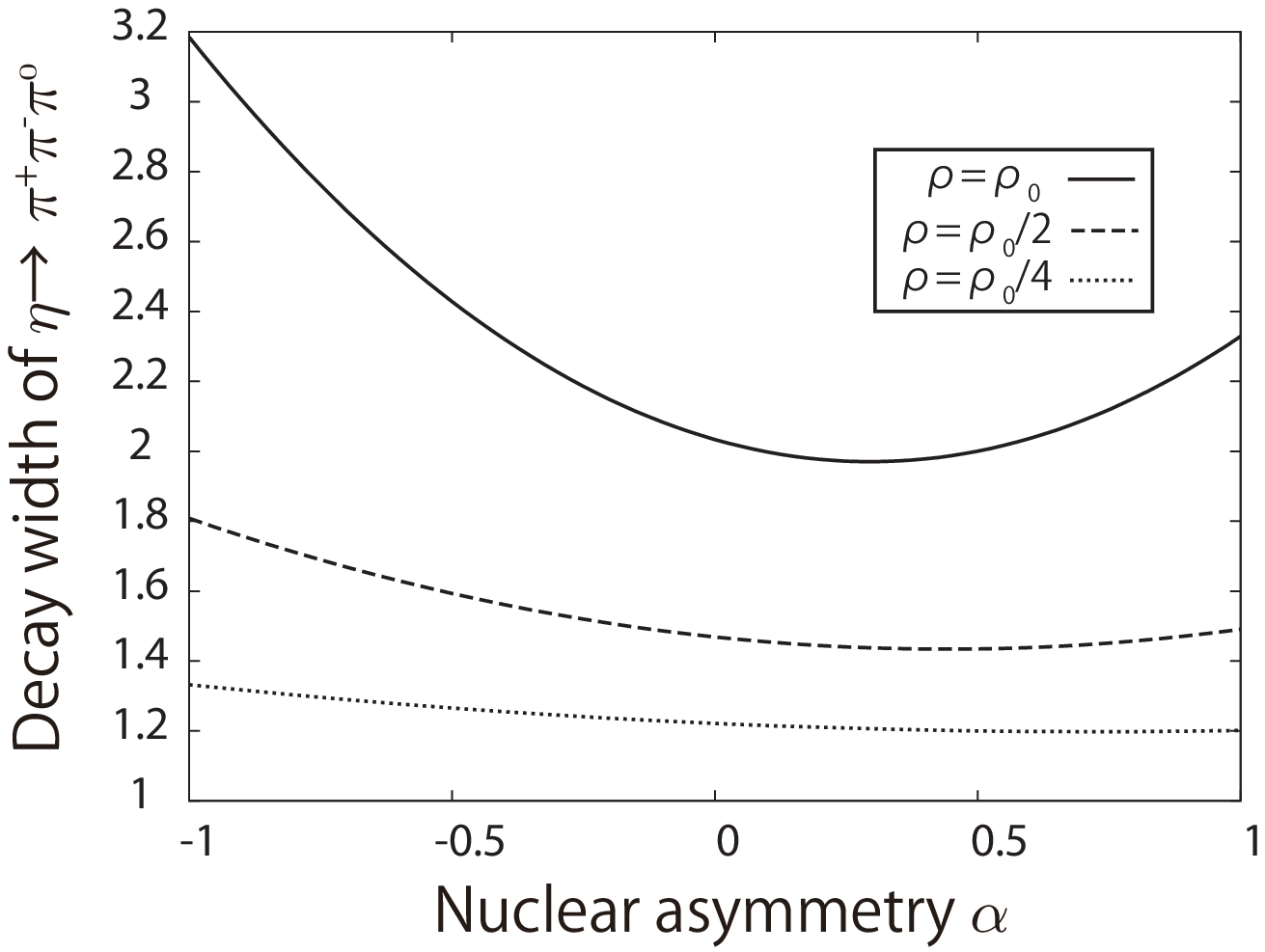}
 \caption{The  nuclear asymmetry $\alpha$
 dependence of the $\eta$ decay width into $\ppp$ with  some fixed
 $\rho$.
 The solid, dashed, and dotted lines are the decay widths at
 $\rho=\rho_0$, $\rho_0/2$, and $\rho_0/4$, respectively. 
 The decay width is normalized with the value at $\rho$ and $\delta\rho=0$.}
 \label{drho} 
\end{minipage}
\hfill
\begin{minipage}[t]{0.475\hsize}
 \centering
 \includegraphics[width=6cm]{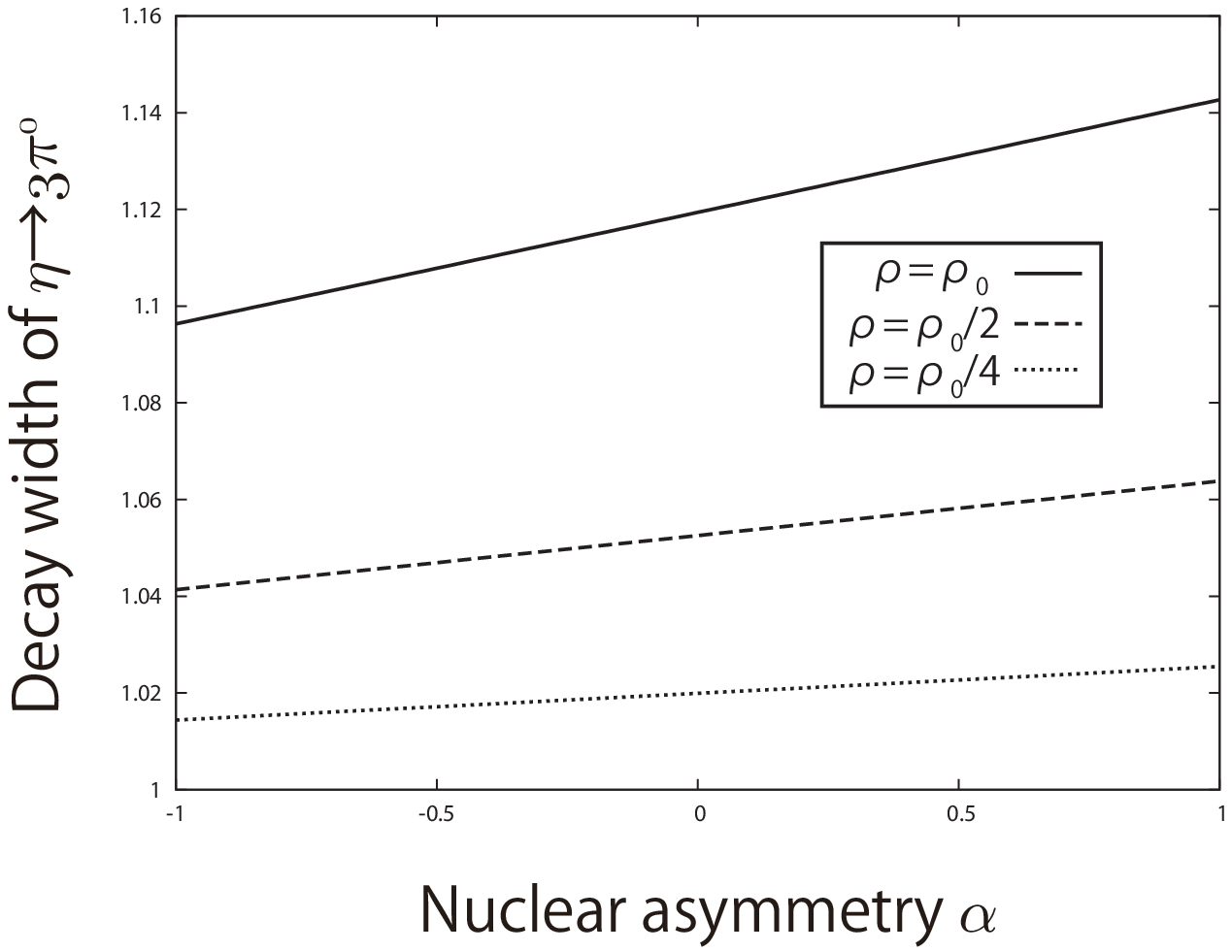}
 \caption{The  nuclear asymmetry $\alpha$
 dependence of the $\eta$ decay width into three $\pi^0$ with
 some fixed $\rho$.
 The solid, dashed, and dotted lines are the decay widths at $\rho=\rho_0$,
 $\rho_0/2$, and $\rho_0/4$, respectively.
 The decay width is normalized with the value at $\rho$ and $\delta\rho=0$.}
 \label{drho2} 
\end{minipage}
\end{figure}

First, we discuss the $\eta-\ppp$ decay width.
From Fig.~\ref{widthppp}, one finds that the width is large in the
higher-density region (upper right of the figure), and the decay
width is enhanced in the proton-rich region.
We show the $\delta\rho$ dependence of the decay width with some
values of fixed $\rho$ in Fig.~\ref{drho}.
The parabolic-shape dependence on $\delta\rho$ comes from the
$\sin^2\theta$ dependence of the decay width:
The minimum points of the decay width are not located at $\alpha=0$ due to
the explicit symmetry breaking in free space caused by the different
$u,d$ quark masses.
We find that the decay width is enhanced by the total baryon
density $\rho$.

Next, we discuss the $\eta-3\pi^0$ decay.
Figure \ref{width3pi0} is a contour plot of the $\eta-3\pi^0$ decay
in the asymmetric nuclear medium and Fig.~\ref{drho2}
shows the $\delta\rho$ dependence of the $\eta-3\pi^0$
decay width with some fixed values of the total density $\rho$.
The decay width of $\eta$ into 3$\pi^0$ shows an enhancement by the
total baryon density in much the same way as $\eta$ into $\ppp$ decay.
This enhancement is caused by the third term in Eq.~(\ref{m3pi0}).
One finds that the effect of the total baryon density
$\rho$ overwhelms that of the isospin asymmetry $\delta\rho$ on
the $\eta-3\pi^0$ decay. Nonetheless 
the neutron-rich medium enhances the decay width as one can
see in Fig.~\ref{drho2}.
The relative smallness of the effect of the isospin asymmetry
$\delta\rho$ on the decay 
in comparison with that on the $\ppp$ decay can be attributed to the fact
that the mixing angle dependence of the decay amplitude is suppressed by
the crossing symmetry.
 
\begin{figure}[t]
\begin{minipage}[t]{0.475\hsize}
 \centering
 \includegraphics[width=6cm]{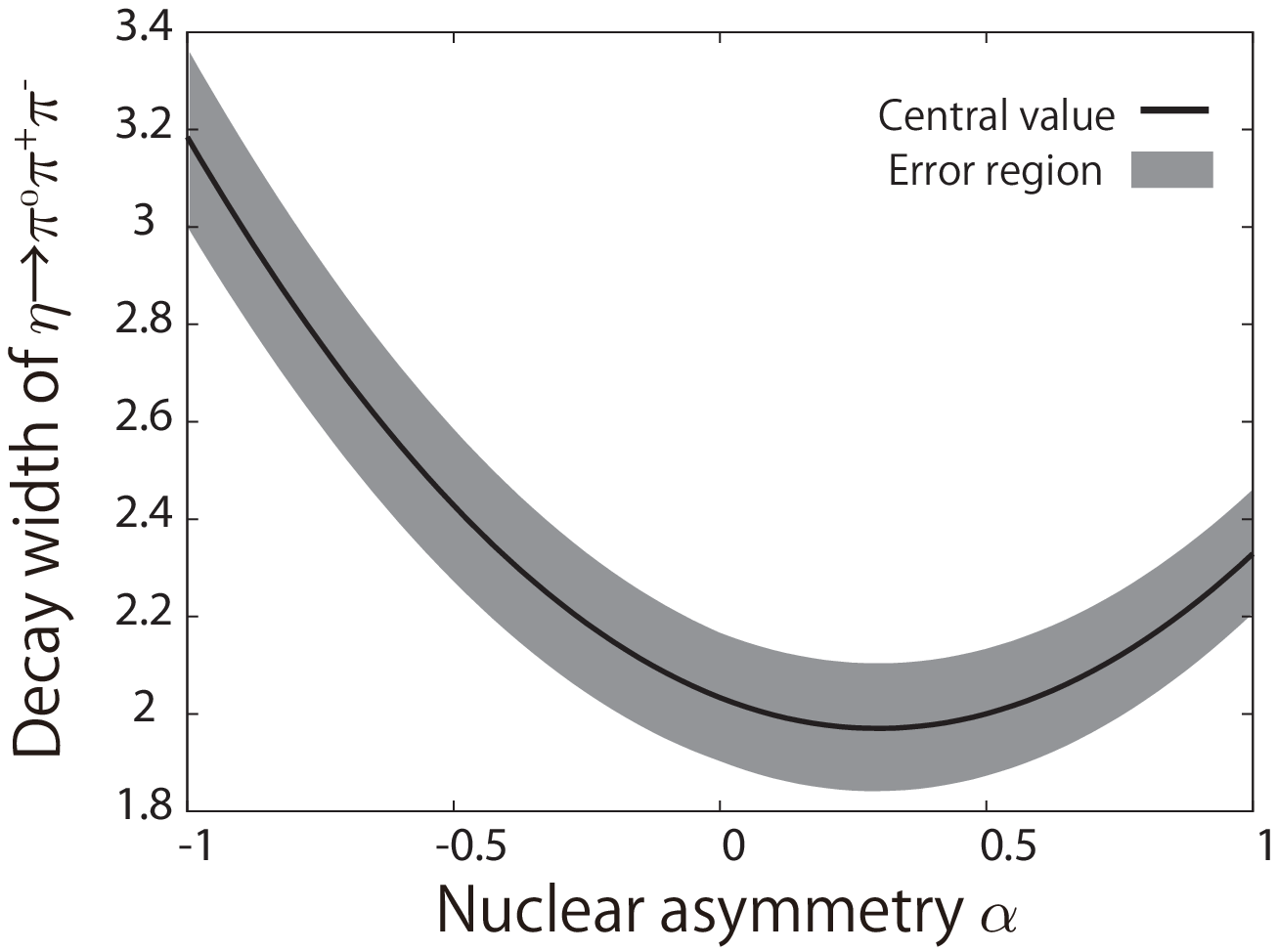}
 \caption{
 The uncertainty of the $\eta$ decay width into $\ppp$ at normal
 nuclear density.
 The solid line corresponds to the decay width with the LECs at the
 central values.
 The shaded area represents the uncertainty of the width due to that of
 the LECs.
 The others are the same as those of Fig.~\ref{drho}.}
 \label{ppp_error} 
\end{minipage}
\hfill
\begin{minipage}[t]{0.475\hsize}
  \centering
  \includegraphics[width=6cm]{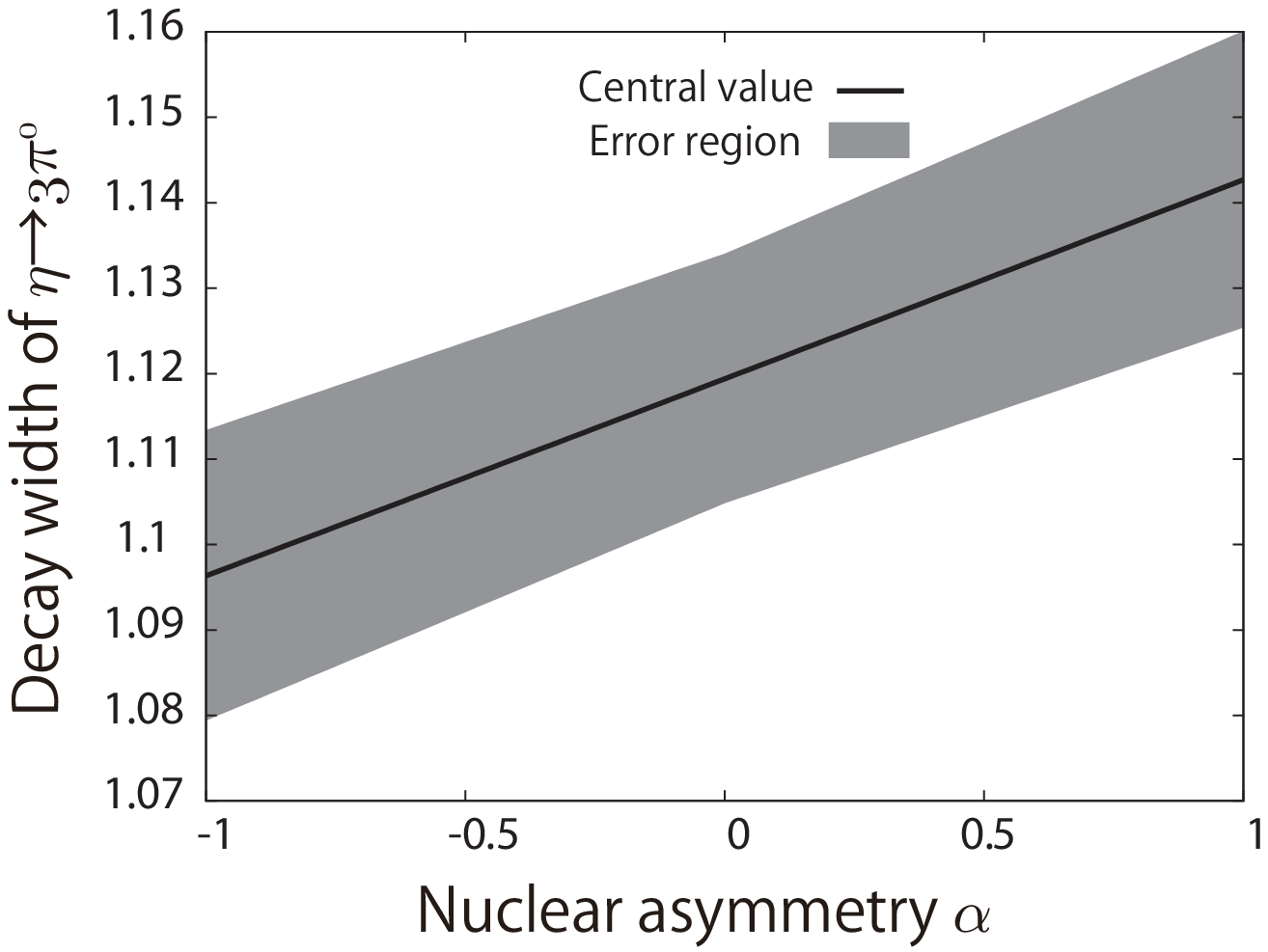}
 \caption{
 The uncertainty of the $\eta$ decay width into $3\pi^0$ 
 due to that of LECs at the normal nuclear density.
 The line and the shaded area are same as those of
 Fig.~\ref{ppp_error}.}
 \label{3p0_error} 
\end{minipage}
\end{figure}

Here, we comment on the uncertainties of the decay widths coming from those
of the LECs.
The uncertainties of the $\eta$ decay width into $\ppp$ and
$3\pi^0$ are both about 10\%.
We present Figs.~\ref{ppp_error} and \ref{3p0_error} to visualize the
uncertainties for normal nuclear densities as an
example:
The solid lines are the central values and the shaded areas show the
uncertainties from the LECs.

Here, we discuss the origin of the enhancement of the decay
widths with the total baryon density.
It is found that the dominant density dependence for both decays comes from 
the term that is proportional to the low-energy constant $c_1$ in
Eqs.~(\ref{mppp2}) and (\ref{m3pi0}) for the $\ppp$ and  3$\pi^0$
decay, respectively.
The parameter $c_1$ is related to $\sigma_{\pi N}$ by
$c_1=-\frac{\sigma_{\pi N}}{4m_\pi^2}$ because we assume that all
the loop corrections of the nucleon mass are renormalized into
LECs \cite{Goda2013}.
Substituting this relation into the second term of Eq.~(\ref{mppp2}),
one finds that the prefactor of the second term reads
\[
-\frac{4c_1m_1^2}{3\sqrt{3}f^4}\rho=\frac{\sigma_{\pi
N}\rho}{m_\pi^2f^2}\frac{m_1^2}{3\sqrt{3}f^2}.
\]
It is noteworthy here that the coefficient $\sigma_{\pi
N}\rho/m_\pi^2f^2$ 
is nothing but the quantity that represents the reduction rate of the
quark condensate or chiral order parameter in
the nuclear medium with the linear density
approximation \cite{Drukarev1991,Brockmann1996};
\begin{align}
 \frac{\delta\left<\bar{q}q\right>}{\left<\bar{q}q\right>_{\rho=0}}
\equiv\frac{\left<\bar{q}q\right>_{\rho=0}-
\left<\bar{q}q\right>_\rho}{\left<\bar{q}q\right>_{\rho=0}}=\frac{\sigma_{\pi N}}{m_\pi^2f^2}\rho,\label{dqq}
\end{align}
where $\left<\bar{q}q\right>_{\rho=0}$ and
$\left<\bar{q}q\right>_\rho$ are the quark condensates at $\rho=0$ and
non-zero, respectively.
Thus, one should be able to rewrite the decay amplitude given in
Eq.~(\ref{mppp2}) (Eq.~(\ref{m3pi0})) for the $\ppp$ (3$\pi^0$) decay
in terms of the reduction of the chiral order parameter
$\delta\left<\bar{q}q\right>$.
Here, we rewrite the
$\mathcal{M}_{\eta\rightarrow\ppp}$ presented in Eq.~(\ref{mppp2}) in terms
of the renormalized pion decay constant $f^\ast$:
\begin{align}
 &\mathcal{M}_{\eta\rightarrow\ppp}=-\frac{m_1^2}{3\sqrt{3}f^2}\left(1+\frac{\sigma_{\pi
 N}}{m_\pi^2f^2}\rho\right)\left(1+\frac{3(s-s_0)}{m_\eta^2-m_{\pi^0}^2}\right)+\sin\theta^{(0)}\mathcal{M}_{\eta\rightarrow\ppp}^{\rm(4)vac}\notag\\
&\hspace{30pt}+\left\{-\frac{s-s_0}{f^2}\frac{1}{m_{\eta}^2-m_{\pi^0}^2}\left(\frac{g_A^2m_\eta^2}{4\sqrt{3}f^2}+\frac{2c_5m_\pi^2}{\sqrt{3}f^2}\right)+\frac{g_A^2}{48\sqrt{3}f^4}(m_\eta-E_{\pi^0})-\frac{2c_5m_\pi^2}{3\sqrt{3}f^4}\right\}\delta\rho\notag\\
&\hspace{20pt}=-\frac{m_1^2}{3\sqrt{3}f^{\ast 2}}\left(1+\frac{3(s-s_0)}{m_\eta^2-m_{\pi^0}^2}\right)+\sin\theta^{(0)}\mathcal{M}_{\eta\rightarrow\ppp}^{\rm(4)vac}\notag\\
&\hspace{30pt}+\left\{-\frac{s-s_0}{f^2}\frac{1}{m_{\eta}^2-m_{\pi^0}^2}\left(\frac{g_A^2m_\eta^2}{4\sqrt{3}f^2}+\frac{2c_5m_\pi^2}{\sqrt{3}f^2}\right)+\frac{g_A^2}{48\sqrt{3}f^4}(m_\eta-E_{\pi^0})-\frac{2c_5m_\pi^2}{3\sqrt{3}f^4}\right\}\delta\rho,\label{mppp3}
\end{align}
where $f^{\ast 2}=f^2\left(1-\frac{\sigma_{\pi N}}{f^2 m_\pi^2}\rho\right).$
From the first to the second line, we have regarded $\sigma_{\pi
N}\rho/m_\pi^2f^2$ as a small quantity, 
and thus the $\rho$ dependence is absorbed into $f^{\ast 2}$.
As one can see in Eq.~(\ref{mppp3}), the total baryon density dependence
of the decay amplitude of the $\eta\rightarrow\ppp$
process can be renormalized into the density dependence of the pion
decay constant\footnote{We note that Eq.~(\ref{dqq}) can also be
rewritten in terms of $f^{\ast 2}$
under the assumption of the smallness of the  change in the pion mass in
the nuclear medium and the in-medium Gell-Mann--Oakes--Renner relation
\cite{Jido2008}:
\begin{align}
 \frac{f_\pi^{\ast 2}m_\pi^{\ast
 2}}{f_\pi^2m_\pi^2}=\frac{\left<\bar{q}q\right>_\rho}{\left<\bar{q}q\right>_{\rho=0}}.
\end{align}
}.
Thus, one may say that the enhancement of the 3$\pi$ decay width 
originates from the chiral restoration in the nuclear medium, although
more detailed analysis is necessary to
establish the relevance of the partial restoration of the chiral
symmetry in the enhancement of the 
3$\pi$ decay of $\eta$ at finite baryon density.
Nevertheless, it is worth mentioning that 
a similar mechanism has been identified as being responsible 
for an enhancement of the $\pi\pi$  cross section near the 2${\pi}$
threshold in the $\sigma$ meson channel in nuclear matter by Jido,
Hatsuda, and one of the present authors \cite{Jido2000}, where
it is found that the reduction of the chiral condensate implying
the partial  restoration of chiral symmetry in nuclear matter is
responsible for this enhancement.
Furthermore, they clarified
that a 4$\pi$-nucleon vertex shown in Fig.~5 in
Ref.~\cite{Jido2000}, which has
the same structure as diagram ($g$) in Fig.~\ref{diagrams} in the
present article, is responsible for the enhancement.

\section{Brief summary and concluding remarks\label{conc}}

In this paper, we have studied the $\eta-\pi^0$ mixing angle defined
in Eq.(\ref{defma})
and the $\eta$ decay into 3$\pi$ in the  nuclear medium with varying
isospin asymmetry on the basis of the in-medium chiral effective theory,
where the Fermi momentum $k_f$ as well as 
the pseudoscalar meson masses and momenta are treated as small
expansion parameters.
We have found that both the quantities are significantly 
modified in the nuclear medium by the asymmetry
$\delta\rho=\rho_n-\rho_p$ and the total baryon
densities $\rho=\rho_n+\rho_p$, although the densities affect both the
quantities in different manners:
the mixing angle increases along with the asymmetric density
$\delta \rho$,
but decreases along with the increase of the total baryon density.
In terms of the $\alpha$ dependence of the mixing angle,
the mixing angle increases along with $\alpha$, the
slope of which is greater for larger total baryon density.
The increase of the mixing angle means that the physical $\eta$ meson has 
a greater $\pi_3$ component, the isospin eigenstate of the neutral pion.
It turns out that the total and asymmetric densities tend to
increase the decay width of $\eta$ into $\ppp$ in an additive way,
although the total density dependence overwhelms that of the isospin.
For example, the width is enhanced by a factor of $2$ to $3$ at
$\rho=\rho_0$ depending on the isospin asymmetry $\alpha$. 

The  enhancement of the width due to the isospin asymmetry 
is traced back to the increase of the $\pi_3$ component in the physical
$\eta$ state.

The $\eta-3\pi^0$ decay width in the nuclear medium 
is enhanced by the total baryon density $\rho$ as 
 in the case of  $\eta\rightarrow\pi^0\pi^+\pi^-$ decay, although
the enhancement is relatively smaller than the latter case.
This is because the mixing angle dependence of the $\eta-3\pi^0$ decay
amplitude is suppressed by the symmetrization of the 
three $\pi^0$ in the final state.
The neutron-rich asymmetric medium slightly enhances the decay width in
the $\eta-3\pi^0$ case.

The $\rho$-dependent parts of the decay amplitudes of the $\eta$ to 3$\pi$
decay given in Eqs.~(\ref{mppp2}) and (\ref{m3pi0}) 
are both proportional to the low-energy constant $c_1$, which is
in turn related to the $\pi-N$ sigma term $\sigma_{\pi N}$.
Thus, we find that the effect of the total baryon density can be
rewritten in terms of the change of the quark condensate in the nuclear
medium.
This suggests that the enhancement of the decay rate may result from the
partial restoration of chiral symmetry in the nuclear medium.
Further analyses are, however, necessary to elucidate the underlying
mechanism  of the enhancement and its possible relationship with the
chiral restoration in the nuclear medium.

It may be possible to observe the modification with the nucleus target
experiment with large isospin asymmetry $(N-Z)/A$ at facilities such
as, e.g., SPring-8 and J-PARC in Japan or FAIR, COSY, and MAMI in
Germany.

In our calculation, the decay widths in free space reproduce
the experimental data fairly well:
The values are roughly 60\% in the $\eta\rightarrow\ppp$ case
and 70\% in the $\eta\rightarrow 3\pi^0$ case compared with the
experimental data \cite{Beringer2012}.
To reproduce the experimental data in free space more precisely,
the higher-order terms in the chiral perturbation should be
included \cite{Bijnens2007}.
Some of them may be resummed into the final-state interaction of $\pi$,
more precisely, the s-wave $\pi^+\pi^-$ correlations, which may cause
the $\sigma$ resonance as mentioned in Sect.~\ref{intro}.

The nuclear medium would affect the spectral properties of the $\sigma$
mode through chiral restoration
\cite{Hatsuda1994,Kunihiro:1995rb,Hatsuda1999,Jido2000,Hyodo2010} (see
also, e.g., Refs.~\cite{Hatsuda1994,Brown1996}), and
thus a full account of the medium effect on the $\sigma$ mode including
its possible softening may have a significant impact on the $\eta$ decay
in the nuclear medium.

Our calculations do not
take the effect of the $N^{\ast}$(1535) resonance into account as
mentioned in Sect.~\ref{secma}.
It is known \cite{Hirenzaki2010} (see also, e.g., Ref.~\cite{Kelkar2013}
for a recent review) that the coupling with the resonance modifies the
in-medium self-energy or the optical potential of $\eta$:
It is suggested that the $\eta$-$N$ interaction is attractive,
and the attraction might lead to a reduction of the $\eta$ mass, say of
$50$ MeV order, in the nuclear medium.
Furthermore, there arises an induced $\eta-\pi^0$ coupling through the
 $N^{\ast}$-nucleon hole excitations in the nuclear medium. 
The formula of the $\eta-\pi^0$ mixing angle in Eq.~(\ref{defma})
tells us that both the reduction of the $\eta$ mass and the additional 
$\eta-\pi^0$ coupling would cause an additional $\rho$ dependence but
not the $\delta \rho$ dependence of the mixing angle and hence an
enhancement of the  $\ppp$ decay width.
On the other hand, the 3$\pi^0$ decay is independent of the mixing
angle, as shown in Eq.~(\ref{m3pi0}), so the effect of the resonance on the
3$\pi^0$ decay width should be small.
We hope that we can report on a quantitative analysis of such an
additional  density dependence coming from the coupling with $N^{\ast}$
of the mixing angle and the decay width of the $\eta$ in the near future.

Recently, the possible effects of an external magnetic field on hadron
properties have been the focus of intensive studies;
the relevant physical quantities include mass spectra of
light hadrons\cite{Meissner2007,Andersen2012}
and pseudoscalar$-$vector mixing rates in heavy quarkonia,
as well as their mass shifts \cite{Alford2013,Cho2014}.
We note that a strong magnetic field could also be a possible source of
the isospin symmetry breaking due to the difference in the
electromagnetic charges of the $u$ and $d$ quarks, and thus 
an external magnetic field may change the hadron properties, as we have
shown in the present work.
This subject is left for future study.

\section*{Acknowledgments}
S.~S. is a JSPS fellow and appreciates the support of a JSPS
Grant-in-Aid (No.~25-1879).
T.~K. was partially supported by a Grant-in-Aid for Scientific Research
from the Ministry of Education, Culture, Sports, Science and Technology
(MEXT) of Japan 
(Nos.~24340054 and 24540271), by  the Core Stage Back UP program 
in Kyoto University, and by the Yukawa
International Program for Quark$-$Hadron Sciences.

\appendix
\section{Meson$-$Baryon vertices\label{vertices}}
In this appendix, we present the meson$-$baryon vertices used in the
calculation in the text.

We first recall the following formula for the derivative of the
exponential operator:
\begin{align}
 e^{-iA(t)}\frac{d}{dt}e^{iA(t)}=i\frac{dA}{dt}+\frac{1}{2!}[A,\frac{dA}{dt}]-\frac{i}{3!}[A,[A,\frac{dA}{dt}]]-\frac{1}{4!}[A,[A,[A,\frac{dA}{dt}]]]+\cdots.
\end{align}
In the Lagrangian given in Eq.~(\ref{lag}), $u^\dagger\partial_\mu u$ and
$u\partial_\mu u^\dagger$ appear.
With the definition of $u=e^{i\pi/2f}$, $u^\dagger\partial_\mu u$ can be
expanded as
\begin{align}
u^\dagger\partial_\mu u= e^{-i\pi/2f}\partial_\mu
 e^{i\pi/2f}=&\frac{i}{2f}\partial_\mu\pi+\frac{1}{2!(2f)^2}[\pi,\partial_\mu\pi]-\frac{i}{3!(2f)^3}[\pi,[\pi,\partial_\mu\pi]]\notag\\
&-\frac{1}{4!(2f)^4}[\pi,[\pi,[\pi,\partial_\mu\pi]]]+\cdots.\label{formula}
\end{align}
In the case of $u\partial_\mu u^\dagger$, the terms with odd powers of
the $\pi$ field change their signs.

\subsection{$\eta\bar{N}N$ and $\pi^0\bar{N}N$ vertices}
With the definitions of $\Gamma_\mu$ and $u$, the $\eta\bar{N}N$ and
$\pi^0\bar{N}N$ vertices from $\mathcal{L}^{(1)}_{\pi N}$
given in Eq.~(\ref{LpiN1}) are given as 
\begin{align}
 i\bar{N}\frac{g_A}{2}\gamma^\mu\gamma_5i\partial_\mu\left(\frac{i\pi}{2f}-\frac{-i\pi}{2f}\right)N=\bar{N}\left(\frac{ig_A}{2f}\gamma_5\gamma^\mu\partial_\mu\pi\right)N,
\end{align}
so the $\eta\bar{N}N$ and $\pi^0\bar{N}N$ vertices, $g_{\eta\bar{N} N}$ and
$g_{\pi^0 \bar{N}N}$, are given as
\begin{align}
g_{\eta \bar{N}N}=&-\frac{g_A}{2f\sqrt{3}}\gamma_5\Slash{k}{\bf 1}\label{genn}\\
g_{\pi^0 \bar{N}N}=&-\frac{g_A}{2f}\gamma_5\Slash{k}\tau_3.\label{gpnn}
\end{align}
Here, the $2\times 2$ matrices $\tau_3$ and {\bf 1} operate on the nucleon
doublet field.

\subsection{$\eta\pi^0\bar{N}N$ vertex}
\subsubsection{The vertex from $\mathcal{L}_{\pi N}^{(1)}$\label{secepnn1}}
The $\eta\pi^0\bar{N}N$ vertex in the leading order comes from
$\mathcal{L}_{\pi N}^{(1)}$ presented in Eq.~(\ref{LpiN1}).
The quadratic terms are induced from the covariant derivative
defined by Eq.~(\ref{cov}).
Using Eq.~(\ref{formula}) and leaving the quadratic term
of $\pi$, we obtain the $\pi^a\pi^b\bar{N}N$ vertex as
\begin{align}
i\bar{N}\gamma^\mu\frac{i}{2}(u^\dagger\partial_\mu u+u\partial_\mu
 u^\dagger)N=i\bar{N}\gamma^\mu\frac{i}{2!(2f)^2}[\pi,\partial_\mu\pi]N.
\end{align}
In the $\eta\pi^0\bar{N}N$ case, note that $\lambda^3$ and
$\lambda^8$ commute.
Thus, we see that the $\eta\pi^0\bar{N}N$ vertex from $\mathcal{L}_{\pi
N}^{(1)}$ vanishes.

\subsubsection{The vertex from the $c_1$ term\label{secepnnc1}}
The $\eta\pi^0\bar{N}N$ vertex with $c_1$ is given as
\begin{align}
 &\ \ ic_1\left<u^\dagger\chi u^\dagger+u\chi
 u\right>\bar{N}N=ic_1\left<\left(\chi\frac{1}{2!}\left(\frac{-i\pi}{2f}\right)^2+\left(\frac{-i\pi}{2f}\right)\chi\left(\frac{-i\pi}{2f}\right)+\frac{1}{2!}\left(\frac{-i\pi}{2f}\right)^2\chi\right)\right.\notag\\
&\ \ \left.+\left(\chi^\dagger\frac{1}{2!}\left(\frac{i\pi}{2f}\right)^2+\left(\frac{i\pi}{2f}\right)\chi^\dagger\left(\frac{i\pi}{2f}\right)+\frac{1}{2!}\left(\frac{i\pi}{2f}\right)^2\chi^\dagger\right)\right>\bar{N}N\notag\\
 &=-i\frac{8B_0c_1}{4f^2}\left<\mathcal{M}\pi^2\right>\bar{N}N\notag\\
&=-\frac{2B_0c_1}{f^2}\left<
\begin{pmatrix}
 m_u&\\
 &m_d 
\end{pmatrix}
\begin{pmatrix}
 \pi^0+\eta/\sqrt{3}&\sqrt{2}\pi^-\\
 \sqrt{2}\pi^+&-\pi^0+\eta/\sqrt{3} 
\end{pmatrix}^2
\right> \bar{N}N\notag\\
 &=-i\frac{2B_0c_1}{f^2}\left<
\begin{pmatrix}
 m_u&\\
 &m_d 
\end{pmatrix}
\begin{pmatrix}
 (\pi^0+\eta/\sqrt{3})^2+2\pi^+\pi^-&2\sqrt{2/3}\pi^-\eta\\
 2\sqrt{2/3}\pi^+\eta&(-\pi^0+\eta/\sqrt{3})^2+2\pi^+\pi^- 
\end{pmatrix}
\right> \bar{N}N.
\end{align}
Thus, the $\eta\pi^0\bar{N}N$ vertex from the $c_1$ term is given as
\begin{align}
 g_{\eta\pi^0\bar{N}N}^{(c_1)}=-i\frac{4B_0c_1}{\sqrt{3}f^2}(m_u-m_d)=i\frac{4c_1m_1^2}{\sqrt{3}f^2}.\label{gepnnc1}
\end{align}

\subsubsection{The vertex from the $c_5$ term\label{gepnnc5}}
The $\eta\pi^0\bar{N}N$ vertex from the $c_5$ vertex is obtained as
\begin{align}
 &ic_5\left<\bar{N}\left(\chi\frac{1}{2!}\left(\frac{-i\pi}{2f}\right)^2+\left(\frac{-i\pi}{2f}\right)\chi\left(\frac{-i\pi}{2f}\right)+\frac{1}{2!}\left(\frac{-i\pi}{2f}\right)^2\chi\right)N\right.\notag\\
&\ \
 \left.+\bar{N}\left(\chi^\dagger\frac{1}{2!}\left(\frac{i\pi}{2f}\right)^2+\left(\frac{i\pi}{2f}\right)\chi^\dagger\left(\frac{i\pi}{2f}\right)+\frac{1}{2!}\left(\frac{i\pi}{2f}\right)^2\chi^\dagger\right)N\right>-i\frac{c_5}{2}\frac{4m_1^2}{\sqrt{3}f^2}\bar{N}N\notag\\
 =&-i\frac{2B_0c_5}{8f^2}\left<\bar{N}\left(\mathcal{M}\pi^2+2\pi\mathcal{M}\pi+\pi^2\mathcal{M}\right)N\right>
 -i\frac{c_5}{2}\frac{4m_1^2}{\sqrt{3}f^2}\bar{N}N
\end{align}
The relevant terms to $\eta\pi^0\bar{N}N$ are
\begin{align}
=&-i\frac{2B_0c_5}{f^2}\bar{N}
\begin{pmatrix}
 m_u&\\
&m_d
\end{pmatrix}
\begin{pmatrix}
 2\pi^0\eta/\sqrt{3}&\\
 &-2\pi^0\eta/\sqrt{3} 
\end{pmatrix}N-i\frac{c_5}{2}\frac{4m_1^2}{\sqrt{3}f^2}\bar{N}N\notag\\
 =&-i\frac{2B_0c_5}{f^2}\left((\bar{p}p)m_u(2\pi^0\eta/\sqrt{3})-(\bar{n}n)m_d(2\pi^0\eta/\sqrt{3})\right)-i\frac{2c_5m_1^2}{\sqrt{3}f^2}(\bar{p}p+\bar{n}n)
\end{align}
Thus, the $\eta\pi^0\bar{N}N$ vertex from the $c_5$ term
is given by
\begin{align}
 g_{\eta\pi^0\bar{N}N}^{(c_5)}=-i\frac{4B_0c_5}{\sqrt{3}f^2}\text{diag}(m_u,-m_d)-i\frac{2c_5m_1^2}{\sqrt{3}}{\bf 1}.\label{gepnnc5}
\end{align}

\subsection{$\ppp\bar{N}N$ vertex\label{secmmmnn}}
The $\ppp\bar{N}N$ vertex from $\mathcal{L}_{\pi N}^{(1)}$ is given as
\begin{align}
&i\bar{N}\left(\frac{g_A}{2}\gamma^\mu\gamma_5i(u^\dagger\partial_\mu
 u-u\partial_\mu u^\dagger)\right) =i\bar{N}\left(\frac{g_A}{2}\gamma^\mu\gamma_5(2i)(-\frac{i}{3!(2f)^3})[\pi,[\pi,\partial_\mu\pi]]\right)N\notag\\
 =&i\bar{N}\left(\frac{g_A}{2}\gamma^\mu\gamma_5(2i)(-\frac{i}{3!(2f)^3})[\lambda^a,[\lambda^b,\lambda^c]]\pi^a\pi^b\partial_\mu\pi^c\right)N
 =\frac{-g_A}{48f^3}\bar{N}\gamma^\mu\pi^a\pi^b\partial_\mu\pi^c[\lambda^a,[\lambda^b,\lambda^c]]N.\label{mmmbb}
\end{align}
If the $\eta$ field is contained in the $\pi$, the
commutator vanishes because the $\eta$ field commutes
with all the $\pi^{0,\pm}$ fields.
With the Fourier transformation and the formula
$[\lambda^1+\lambda^2,[\lambda^1+\lambda^2,\lambda^3]]=2\lambda^3$, the
$\ppp\bar{N}N$ vertex is obtained as
\begin{align}
 g_{\ppp\bar{N}N}=-\frac{1}{24f^3}\gamma_5\gamma_\mu(2p^\mu_0-p_+^\mu-p_-^\mu).\label{gpppnn}
\end{align}
Because $a,b,c=\pi^0$ in Eq.~(\ref{mmmbb}), the $3\pi^0$ vertex
vanishes.

\subsection{$\eta\ppp\bar{N}N$ and $\eta3\pi^0\bar{N}N$
vertices\label{secmmmmnn}}
\subsubsection{The vertex from $\mathcal{L}_{\pi N}^{(1)}$\label{secmmmmnn1}}
The four mesons and the $\bar{N}N$ vertex from $\mathcal{L}_{\pi
N}^{(1)}$ are given as
\begin{align}
 &i\bar{N}\gamma^\mu\frac{i}{2}\left(u\partial_\mu
 u^\dagger+u^\dagger\partial_\mu
 u\right)N=\bar{N}\left(\frac{1}{4!(2f)^4}[\pi,[\pi,[\pi,\partial_\mu\pi]]]\right)N\notag\\
=&\bar{N}\left(\frac{1}{4!(2f)^4}[\lambda^a,[\lambda^b,[\lambda^c,\lambda^d]]]\pi^a\pi^b\pi^c\partial_\mu\pi^d\right)N.
 \label{gepppnn1}
\end{align}
Here, we consider the quartic terms of $\pi$ of Eq.~(\ref{formula})
because they are relevant to the $\eta\ppp\bar{N}N$ and $\eta
3\pi^0\bar{N}N$ vertices.
Because the $\eta$ field is contained, the commutator of the Gell-Mann
matrices vanishes.
Accordingly, the $\eta\ppp\bar{N}N$ and $\eta3\pi^0\bar{N}N$
vertices in $\mathcal{L}_{\pi N}^{(1)}$ vanish. 

\subsubsection{The vertex from the $c_1$ term\label{secmmmmnnc1}}
The $\eta\ppp\bar{N}N$ vertex from the $c_1$ term is given as
\begin{align}
 &ic_1\left<u^\dagger\chi u^\dagger+u\chi^\dagger
 u\right>\bar{N}N\notag\\
=&\frac{ic_1}{(2f)^4}\left<(\chi\frac{1}{4!}\pi^4+\pi\chi\frac{1}{3!}\pi^3+\frac{\pi^2}{2!}\chi\frac{\pi^2}{2!}+\frac{\pi^3}{3!}\chi\pi+\frac{\pi^4}{4!}\pi^4\chi)\right.\notag\\
&\left.+(\chi^\dagger\frac{1}{4!}\pi^4+\pi\chi^\dagger\frac{1}{3!}\pi^3+\frac{\pi^2}{2!}\chi^\dagger\frac{\pi^2}{2!}+\frac{\pi^3}{3!}\chi^\dagger\pi+\frac{\pi^4}{4!}\pi^4\chi^\dagger)\right>\bar{N}N\notag\\
=&\frac{iB_0c_1}{6f^4}\left<\mathcal{M}\pi^4\right>=\frac{B_0c_1}{6f^4}\left<
\begin{pmatrix}
 m_u&0\\
 0&m_d 
\end{pmatrix}
\begin{pmatrix}
 \pi^0+\eta/\sqrt{3}&\sqrt{2}\pi^-\\
 \sqrt{2}\pi^+&-\pi^0+\eta/\sqrt{3} 
\end{pmatrix}^4\right>\bar{N}N\notag\\
=&\frac{iB_0c_1}{6f^4}\left<
\begin{pmatrix}
 m_u&0\\
 0&m_d 
\end{pmatrix}
\begin{pmatrix}
 (\pi^0+\eta/\sqrt{3})^2+2\pi^+\pi^-&2\sqrt{2/3}\pi^-\eta\\
 2\sqrt{2/3}\pi^+\eta&(-\pi^0+\eta/\sqrt{3})^2+2\pi^+\pi^- 
\end{pmatrix}^2\right>\bar{N}N.\label{c1}
\end{align}
Taking only the relevant terms, we obtain the $\eta\ppp\bar{N}N$ vertex
as
\begin{align}
 &\frac{iB_0c_1}{6f^4}\left(m_u\times 8\eta\ppp/\sqrt{3}-m_d\times
 8\eta\ppp/\sqrt{3}\right)\bar{N}N\notag\\
=&\frac{i4c_1B_0}{3\sqrt{3}f^4}(m_u-m_d)\eta\ppp\bar{N}N=-i\frac{4c_1m_1^2}{3\sqrt{3}f^4}\eta\ppp\bar{N}N.
\end{align}
Thus, the $\eta\ppp\bar{N}N$ vertex from the $c_1$ term
is given as
\begin{align}
 g_{\eta\ppp\bar{N}N}^{(c_1)}=-i\frac{4c_1m_1^2}{3\sqrt{3}f^4}.\label{gepppc1}
\end{align}

From Eq.~(\ref{c1}), the $\eta3\pi^0\bar{N}N$ vertex
$g_{\eta3\pi^0\bar{N}N}^{(c_1)}$ is given as
\begin{align}
 g_{\eta3\pi^0\bar{N}N}^{(c_1)}=i\frac{B_0c_1}{6f^4}\frac{4}{\sqrt{3}}(m_u-m_d)=-i\frac{2c_1m_1^2}{3\sqrt{3}f^4}.\label{ge3p0nnc1}
\end{align}

\subsubsection{The vertex from the $c_5$ term\label{mmmmnnc5}}
The $\eta\ppp\bar{N}N$ vertex from the $c_5$ term is
\begin{align}
 &ic_5\bar{N}\left(\frac{1}{(2f)^4}(\chi\frac{1}{4!}\pi^4+\pi\chi\frac{1}{3!}\pi^3+\frac{\pi^2}{2!}\chi\frac{\pi^2}{2!}+\frac{\pi^3}{3!}\chi\pi+\frac{\pi^4}{4!}\pi^4\chi)\right.\notag\\
&\left.+(\chi^\dagger\frac{1}{4!}\pi^4+\pi\chi^\dagger\frac{1}{3!}\pi^3+\frac{\pi^2}{2!}\chi^\dagger\frac{\pi^2}{2!}+\frac{\pi^3}{3!}\chi^\dagger\pi+\frac{\pi^4}{4!}\pi^4\chi^\dagger)\right)N-\frac{ic_5}{2}\left(-\frac{4m_1^2}{3\sqrt{3}f^4}\right)\bar{N}N\notag\\
=&\frac{i4B_0c_5}{4!\cdot(2f)^4}\bar{N}\left(\mathcal{M}\pi^4+4\pi\mathcal{M}\pi^3+6\pi^2\mathcal{M}\pi^2+4\pi^3\mathcal{M}\pi+\pi^4\mathcal{M}\right)N+\frac{i2c_5m_1^2}{3\sqrt{3}f^4}\bar{N}N.\label{c5}
\end{align}
The terms contributing to the $\eta\ppp\bar{N}N$ vertex are given as
\begin{align}
&\frac{B_0c_5}{6f^4}\left(m_u(\bar{p}p)(\frac{8}{\sqrt{3}}\eta\ppp)-m_d(\bar{n}n)(\frac{8}{\sqrt{3}}\eta\ppp)\right)+\frac{i2c_5m_1^2}{3\sqrt{3}f^4}\bar{N}N\notag\\
 =&\frac{i4B_0c_5}{3\sqrt{3}f^4}(m_u\bar{p}p-m_d\bar{n}n)\eta\ppp+\frac{i2c_5m_1^2}{3\sqrt{3}f^4}(\bar{p}p+\bar{n}n)\eta\ppp.
 \end{align}
Thus, the $\eta\ppp\bar{N}N$ vertex is obtained as
\begin{align}
 g_{\eta\ppp\bar{N}N}=i\frac{4B_0c_5}{3\sqrt{3}f^4}\text{diag}(m_u,-m_d)+i\frac{2c_5m_1^2}{3\sqrt{3}f^4}{\bf 1}.\label{gepppnnc1}
\end{align}
From Eq.~(\ref{c5}), the $\eta3\pi^0\bar{N}N$ vertex is given as
\begin{align}
 g_{\eta3\pi^0\bar{N}N}=i\frac{2B_0c_5}{3\sqrt{3}f^4}\text{diag}(m_u,-m_d)+\frac{c_5m_1^2}{3\sqrt{3}f^4}{\bf
 1}.\label{gm3p0nn}
\end{align}

\end{document}